\documentclass[12pt,preprint]{aastex}
\newcommand{\chandra}{{\it Chandra}}
\newcommand{\rosat}{{\it ROSAT}}
\newcommand{\einstein}{{\it Einstein}}
\newcommand{\xmm}{{\it XMM-Newton}}
\newcommand{\hst}{{\it Hubble Space Telescope}}
\newcommand{\hii}{\ion{H}{2}}
\newcommand{\hi}{\ion{H}{1}}
\newcommand{\ha}{H$\alpha$}
\newcommand{\nii}{\ion{N}{2}}
\newcommand{\half}{\small 1/2}

\begin{document}

\title{Supernova Remnants in the Magellanic Clouds V: 
The Complex Interior Structure of the N206 SNR}
\author{R.~M. Williams, Y.-H. Chu,  J.~R. Dickel \& R.~A. Gruendl}
\affil{University of Illinois at Urbana-Champaign, 1002 W. Green St., 
Urbana, IL 61801 USA}
\email{rosanina@astro.uiuc.edu, chu@astro.uiuc.edu, johnd@astro.uiuc.edu, 
gruendl@astro.uiuc.edu}
\author{F.~D. Seward}
\affil{Harvard-Smithsonian Center for Astrophysics, 
60 Garden Street, MS 4, Cambridge, MA 02138 USA}
\email{fds@cfa.harvard.edu}
\author{M.~A. Guerrero\altaffilmark{1}}
\affil{Instituto de Astrof\'isica de Andaluc\'ia, 
Consejo Superior de Investigaciones Cient\'ificas (CSIC), 
Apartado Correos 3004, E-18080 Granada, Spain}
%\altaffiltext{1}{also, University of Illinois at Urbana-Champaign, 1002 
%W. Green St., Urbana, IL 61801 USA}
\email{mar@iaa.es}
\and
\author{G. Hobbs}
\affil{Australia Telescope National Facility, P.O. Box 76, Epping, 
	NSW 1710, Australia}
\email{ghobbs@atnf.csiro.au}

\begin{abstract}
The N206 supernova remnant  (SNR) in the Large Magellanic Cloud (LMC) 
has long been considered a prototypical ``mixed morphology" SNR.  Recent 
observations, however, have added a new twist to this familiar plot:
an elongated, radially-oriented radio feature seen in projection against 
the SNR face.  Utilizing the high resolution and sensitivity available with
the \hst, \chandra, and \xmm, we have obtained optical emission-line 
images and spatially resolved X-ray spectral maps for this intriguing
SNR.  Our findings present the SNR itself as a remnant in the mid to
late stages of its evolution. X-ray emission associated with the radio
``linear feature" strongly suggests it to be a pulsar-wind nebula (PWN). 
A small X-ray knot is discovered at the outer tip of this feature. 
The feature's elongated morphology and the surrounding wedge-shaped
X-ray enhancement strongly suggest a bow-shock PWN structure.
\end{abstract}


\section{Introduction}

The supernova remnant (SNR) B 0532$-$71.0 lies to the north and east of the 
\hii\ region L\ha\ 120-N206\footnote{Catalogue number from \citet{H56}} 
in the Large Magellanic Cloud (LMC). The SNR was first identified in radio 
by \citet{MC73}, who referred to it simply as N206. Although this is more 
properly the name of the larger \hii\ complex, for simplicity we shall use 
the same notation in the following discussion.

\ha\ images of N206 obtained from the Magellanic Cloud Emission-Line Survey 
\citep{SM99} show largely shell-like emission, with complex filamentary 
structure along the limb.  \citet{W+99} noted that these features were in 
sharp contrast to the X-ray emission as observed by \rosat, which shows
the X-ray emission brightening toward the center of the remnant.  
The striking difference between X-ray and optical morphologies led the 
authors to categorize N206 as a ``centrally brightened" SNR, and to
suggest that it might undergo the same physical processes as ``mixed 
morphology" SNRs \citep{RP98}, which have centrally brightened X-ray 
emission and a radio shell. 

\citet{M+80} studied N206 at 5 and 14.7 GHz, and found a spectral index
$\alpha$ (as in $S\sim\nu^{\alpha}$) over the remnant of about $-$0.33.  
This value is fairly flat for a SNR; it is on the borderline between the 
ranges for a pulsar wind nebula (PWN), for which spectral indices tend to 
be flatter than $-$0.4, and for a typical shell-type remnant, for which 
spectral indices tend to be between $-$0.8 and $-$0.3 \citep{T+00}.  Low 
resolution (several-arcminute scales) and interference from the nearby 
\hii\ region made features in the SNR interior difficult to examine.

Of particular interest, therefore, are the observations by \citet{K+02} 
of N206 at 3 and 6 cm wavelengths (8.0 and 4.8 GHz), using the Australia 
Telescope Compact Array (ATCA) with resolutions of 1\farcs8 and 1\farcs1 
(HPBW), respectively.  They saw emission over the face of the entire SNR, 
somewhat brightened toward the limb. The spectral index calculated from 
their findings and those of other radio observations was $-$0.20$\pm$0.07, 
in the normal range for filled-center SNRs.  \citet{K+02} also discerned 
a ``peculiar linear feature" within the SNR: a narrow wedge of radio 
emission aligned in projection with the remnant's center.  A spectral index 
map of the SNR showed this feature to have a similar index to the rest of 
the remnant, $\sim -$0.2.  In the absence of high-resolution X-ray data, 
the authors could not isolate a point source, and therefore could only 
speculate that the feature was produced by ``a low-mass star or compact
object."  Using the ``opening angle" of the linear feature to calculate the 
Mach number of the presumed object's travel, and the length of the linear 
feature, the authors obtained an estimate for the SNR age of 23,000 yr.  

\section{Observations}

\subsection{X-ray Images and Spectroscopy}

X-ray observations of N206 were made with both the \chandra\ and \xmm\ 
observatories. \chandra\ observations used the Advanced CCD Imaging 
Spectrometer (ACIS), primarily the S3 back-illuminated chip. Datasets
were sequence number 500327, observations 3848 (33.1 ks) and 4421 
(32.5 ks).  Data were reduced following procedures recommended by the 
Chandra X-ray Center (CXC) and analyzed using the CXC's \textsc{ciao} 
program and \textsc{xspec}. The two datasets were filtered for 
high-background times and poor event grades, resulting in a total 
``good time" interval of 65.6 ks. For each dataset, the 0\farcs5 
pixel randomizations were removed and the ``subpixel resolution" 
technique\footnote{This technique, developed by Koji Mori, allows 
users to improve the spatial resolution by $\sim$10\% 
without decreasing the overall count rates. Documentation at  
\url{http://cxc.harvard.edu/cont-soft/software/subpixel\_resolution.1.4.html}}
was applied, allowing us to more closely examine the structure
on small spatial scales.  

The filtered datasets were then merged, and spectral results extracted 
from the merged file. As spectral analysis required us to choose fairly 
wide regions for reasonable counting statistics, the ``subpixel resolution"
files were not used for this purpose.    A background region was taken 
from an annulus surrounding the SNR, and the spectrum of this background 
region was scaled and subtracted from the source spectra.  Individual
spectra for regions of interest selected from radio and X-ray images,
and the corresponding primary and auxiliary response files, were extracted 
with the \textsc{ciao} psextract script and analyzed in \textsc{xspec}.
Spectra were rebinned by spectral energy to achieve a 
signal-to-noise ratio of 4 in each bin.

We received the pipeline-processed \xmm\ data from the Science Operations 
Centre (SOC). Observations were made simultaneously with multiple \xmm\
instruments;  in this paper we will concentrate on the European Photon 
Imaging Camera (EPIC) MOS and pn detectors. The EPIC-MOS data were taken 
over two intervals, in 2001 November (Observation ID 089210901, 41.4 ks) 
and 2001 December (0089210101, 14.5 ks). The latter observation also 
included an EPIC-pn exposure (12.0 ks). Initial reduction and analysis 
were carried out using the Science Analysis Software (SAS) package 
provided by the SOC, with subsequent spectral analysis in \textsc{xspec}. 

The data were filtered to remove high background times or poor event 
grades, reducing the total ``good" dataset time to 24.2 ks for the first 
observation (EPIC-MOS only) and 12.0 ks for the second observation 
(EPIC-MOS and EPIC-pn).  Images and spectra were then extracted from 
the filtered event files.  Background regions immediately surrounding 
the SNR but free of point sources were used to produce background 
spectra, which were then scaled and subtracted from the source spectra.  
Joint spectral fits were performed with the data from the two EPIC-MOS 
observations. 

\subsection{Optical Emission-line Images and Spectroscopy}

Images of N206 were obtained with the \hst\ using the Wide Field Planetary 
Camera 2 (WFPC2) with the F656N (\ha), F673N ([\ion{S}{2}] $\lambda\lambda$ 
6347, 6371 \AA) and F502N ([\ion{O}{3}] $\lambda\lambda$ 5007 \AA) filters.  
The southwestern side of the SNR was observed in
\ha\ (3$\times$800 s), [\ion{S}{2}] (3$\times$800 s) and [\ion{O}{3}] 
(6$\times$800 s), all at the same pointing and position angle, but the 
northeastern side of the SNR was observed only in \ha\ (3$\times$800 s).
The data were reduced using the \textsc{iraf}\footnote{Image Reduction 
Analysis Facility, maintained by NOAO} program and \textsc{stsdas} package. 
Multiple exposures were combined to remove cosmic-ray events, and the
resulting files were bias-subtracted.  The images were divided by
exposure times to produce count-rate maps, which were then multiplied
by the \textsc{photflam} parameter (provided in the image header) to 
convert these to flux-density maps. We used the \textsc{synphot} task to
determine widths for each filter, and multiplied the flux-density maps
by the filter widths to produce flux maps. The files for 
the WFC and PC were then mosaicked together for the final images.
Images for [\ion{S}{2}]  and \ha\ were clipped at 3$\sigma$ of sky
background to reduce noise, and then divided to produce a calibrated
[\ion{S}{2}]/\ha\ ratio map.

High-dispersion spectra of N206 were obtained with the echelle
spectrograph on the 4 m telescope at Cerro Tololo Inter-American
Observatory (CTIO) from two observing runs, 2000 December 6  and
2004 January 14.  Both runs used the 79 line mm$^{-1}$ echelle
grating in the single-order, long-slit observing configuration,
in which a flat mirror replaced a cross disperser and a post-slit
\ha\ filter ($\lambda_c$ = 6563 \AA, $\Delta \lambda$ = 75 \AA)
was inserted to isolate a single order.  The red long focus camera
and the 2000$\times$2000 SITe2K\#6 CCD were used to record data.
The 24 $\mu$m pixel size corresponds to roughly 0.082 \AA\ along
the dispersion and 0\farcs26 perpendicular to the dispersion.
Limited by the optics of the camera, the useful spatial coverage
is $\sim$3$'$.  The spectral coverage is wide enough to include
both the \ha\ line and the flanking [\nii]~$\lambda\lambda$6548,
6583 lines.  A slit width of 250 $\mu$m (1\farcs64) was used and
the resultant FWHM of the instrumental profile was $13.5\pm0.5$
km s$^{-1}$.  The spectral dispersion was calibrated by a Th-Ar
lamp exposure taken in the beginning of the night, but the
absolute wavelength was calibrated against the geocoronal \ha\
line present in the nebular observations.  The journal of echelle
observations is given in Table~\ref{tab:echobs}.

\subsection{Radio Pulsar Observations}

We undertook new observations using the Parkes 64-m radio telescope at 
the Australia Telescope National Facility.  We observed N206 for 12 hours 
on 2003 September 29 using the central beam of the 20-cm multibeam 
receiver at 1374 MHz on the Parkes telescope with the aim of detecting 
pulsed emission from a small X-ray source. Following the recent 
successful surveys for pulsars in radio nebulae at Parkes 
\citep{C+02a,C+02b}, we recorded total-power signals from 96 frequency 
channels which provide a bandwidth of 288\,MHz for two polarizations 
every 1\,ms. The sensitivity of the system was $\sim 20 \mu$\,Jy which 
is a factor of $\sim$\,2.5 more sensitive than an earlier, large-scale 
survey of the Large Magellanic Cloud \citep{F02}.  Off-line processing 
was carried out to search for any dispersed, periodic signal using 
standard procedures with the \textsc{seek} software.\footnote{Source 
code and documentation are available at 
\url{http://www.jb.man.ac.uk/$\sim$drl/seek}.}  No signal was detected. 

\section{Physical Structure of the SNR}

\begin{figure}
\epsscale{1.0}
\plotone{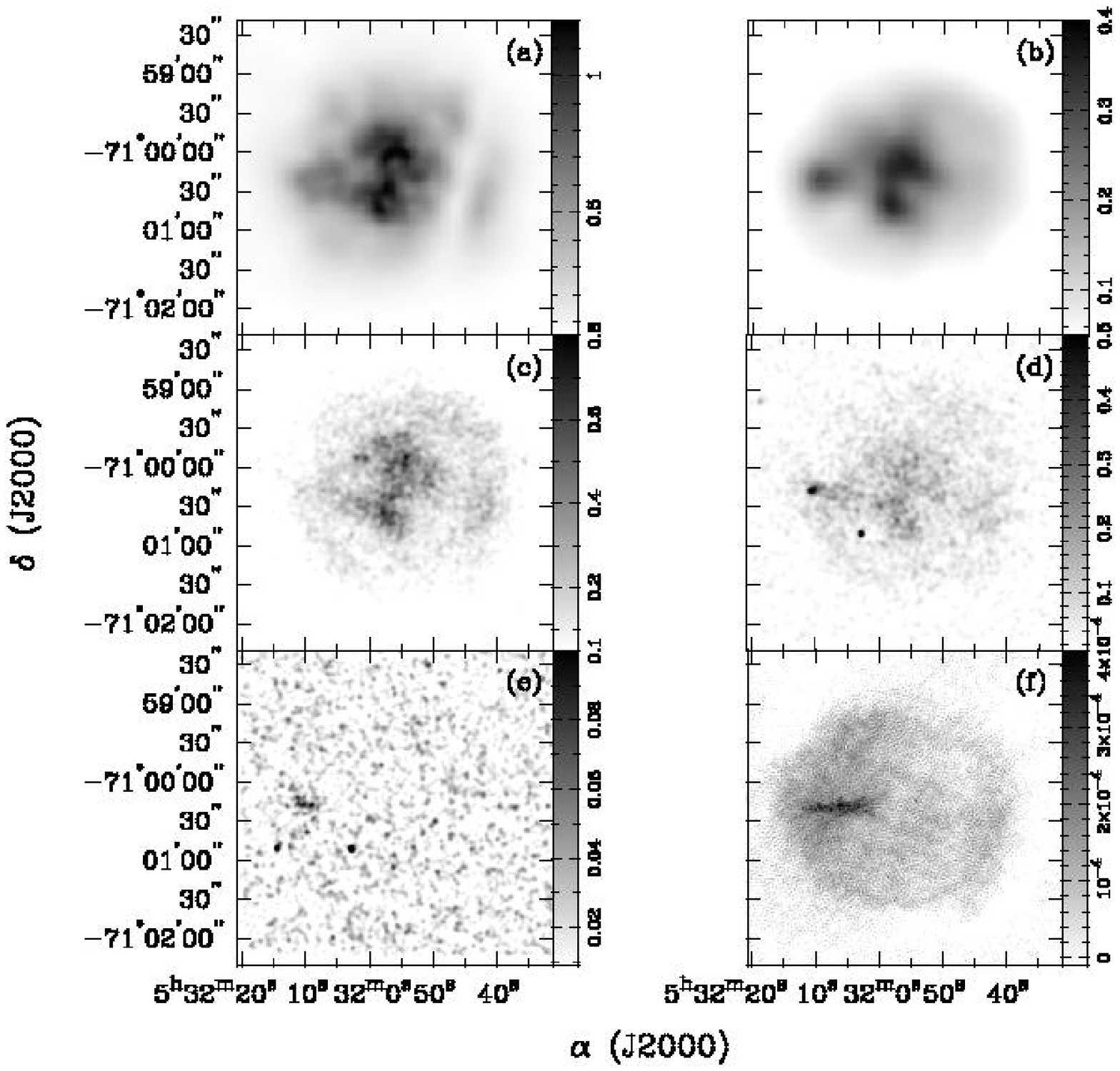}
\caption{Smoothed images, over the same spatial region: 
(a) \xmm\ EPIC-pn 0.2-10.0 keV (b) \xmm\ EPIC-MOS 0.2-10.0 keV
(c) \chandra\ ACIS 0.2-0.9 keV (d) \chandra\ ACIS 0.9-3.0 keV 
(e) \chandra\ ACIS 3.0-8.0 keV (f) ATCA radio 6 cm. \xmm\ 
images are adaptavely smoothed with the same scaling map; 
\chandra\ images are smoothed with a Gausian ($\sigma=3$).}
\label{fig:xmm_acis}
\end{figure}

\begin{figure}
\epsscale{1.0}
\plotone{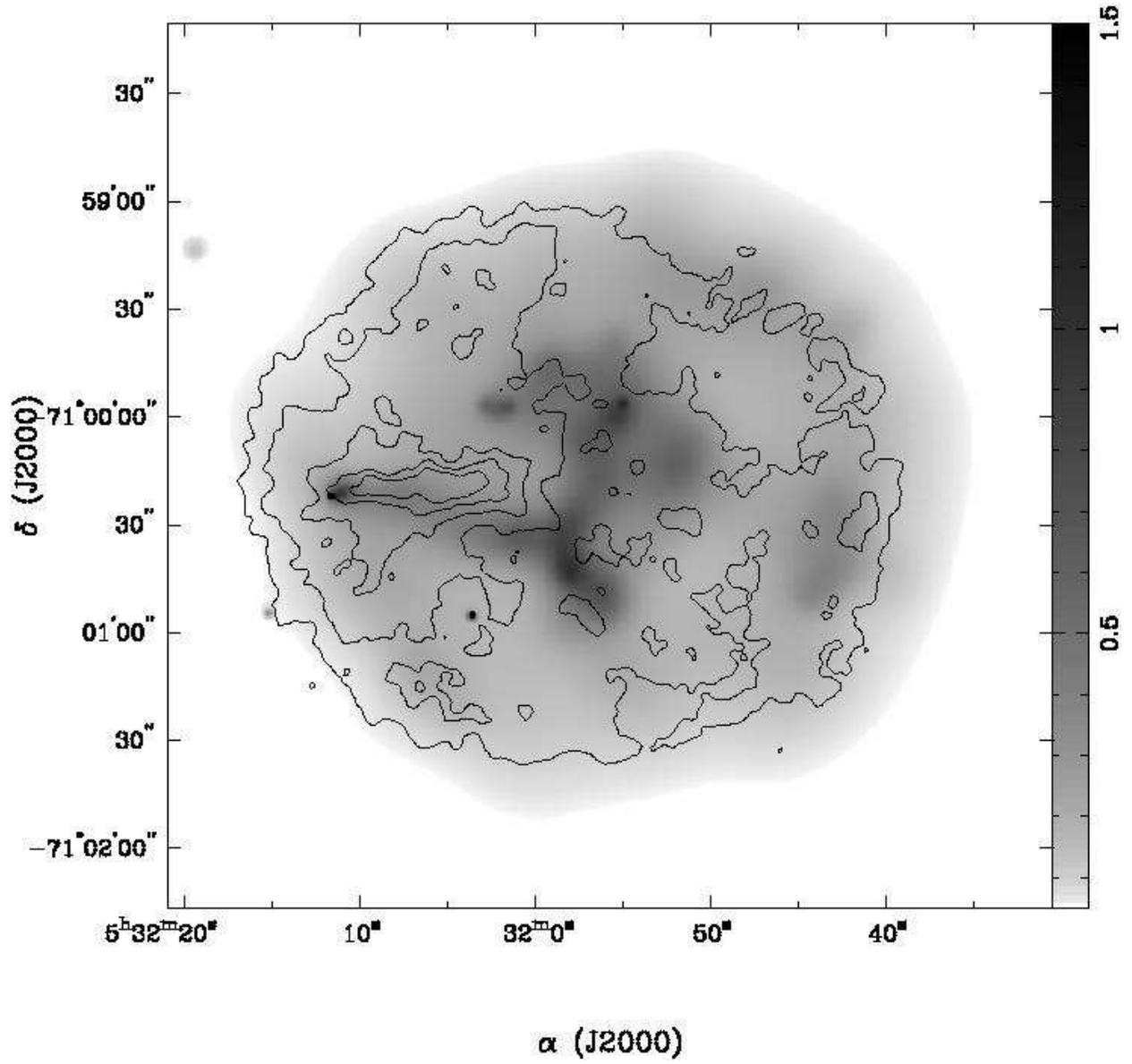}
\caption{Adaptively smoothed Chandra ACIS 0.3-10.0 keV image (grayscale)  
with ATCA 6 cm radio contours overlaid.}
\label{fig:acis_atca}
\end{figure}

The morphology of the SNR in the \xmm\ and \chandra\ observations is 
complex, with particular structures revealing themselves variously under
\xmm's high sensitivity and \chandra's high resolution.  In both cases 
(Fig.~\ref{fig:xmm_acis}) diffuse emission can be seen over the entire 
face of the remnant, all the way out to the rim; and a significant increase 
in X-ray emission appears toward the center.  In the \xmm\ EPIC-pn and
EPIC-MOS observations there is an extension of this central emission that 
corresponds well to the position of the radio ``linear feature".  This 
apparent correlation is confirmed by the \chandra\ observations, in which 
X-ray emission is clearly seen along the length of the linear feature, 
broadening in the same region of the ``wedge" as in radio 
(Fig.~\ref{fig:acis_atca}).

An X-ray point source appears in the \chandra\ observation at 
05$^h$32$^m$03$^s$, $-$71\degr00\arcmin51\arcsec.  No counterpart to
this source is found in the optical or radio data.  A search of the
\textsc{SIMBAD} database shows these coordinates to be within the
error circle for \einstein\ source 2E 0532.6$-$7102 \citep{Mc94}. In
the absence of features indicating an association between this source
and the SNR, we presume it to be a background source; possibly, given
its hard X-ray emission, an active galactic nucleus.  We will not 
include this source in the subsequent discussion.

\begin{figure}
\epsscale{1.0}
\plotone{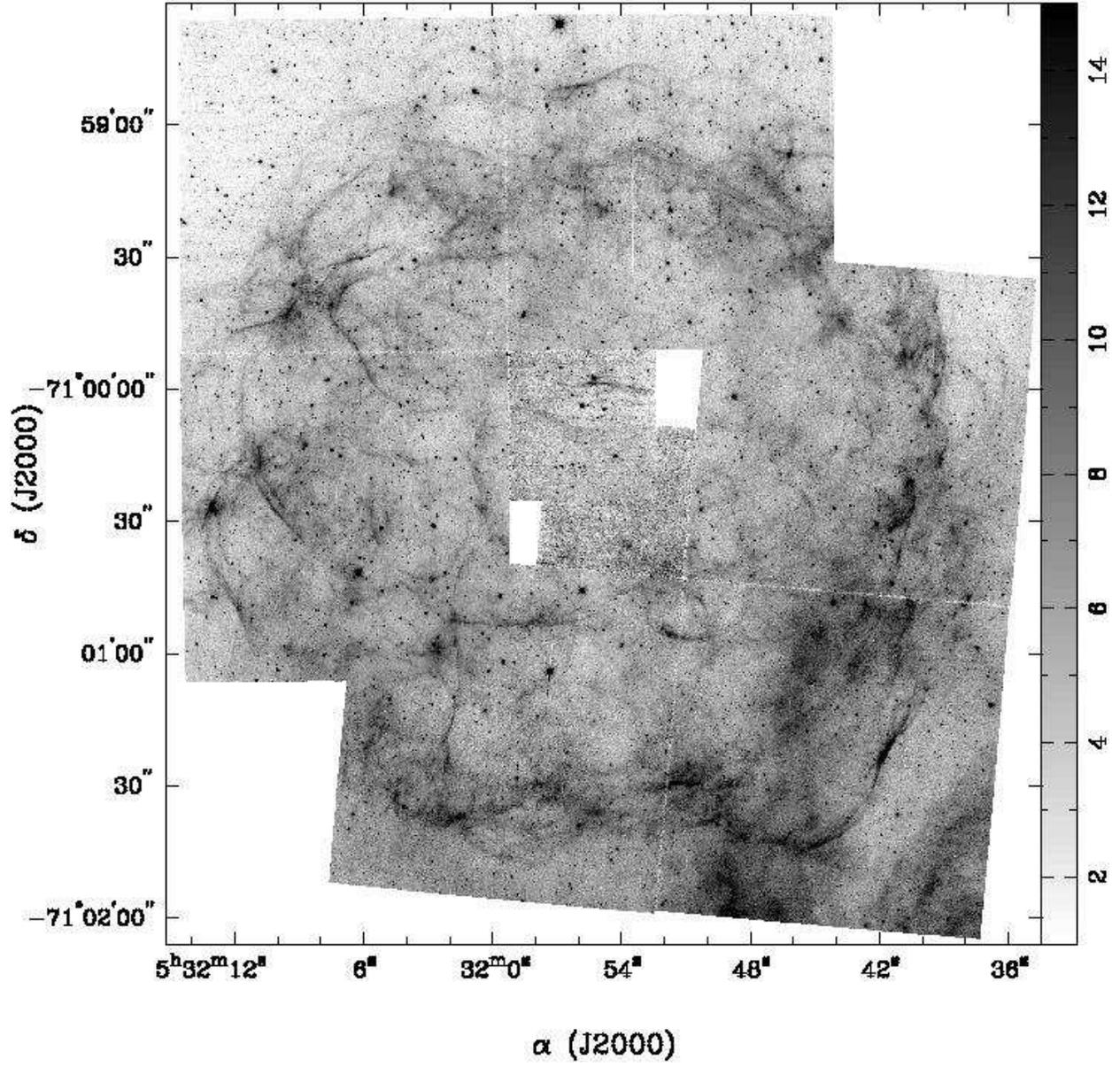}
\caption{Combined HST WFPC2 mosaicked images of N206 in \ha. }
\label{fig:hsthafull}
\end{figure}

The \hst\ WFPC2 \ha\ mosaic of N206 in Figure~\ref{fig:hsthafull}
shows the circular, limb-brightened structure of this SNR. 
An intricate array of loop-like filaments extends from the outermost limb
well within the remnant.  Toward the southwest of the SNR the \ha\ 
emission becomes more prominent; this is unsurprising, as in this region 
the N206 SNR may overlap slightly with the larger N206 \hii\ region.

\begin{figure}
\epsscale{1.0}
\plotone{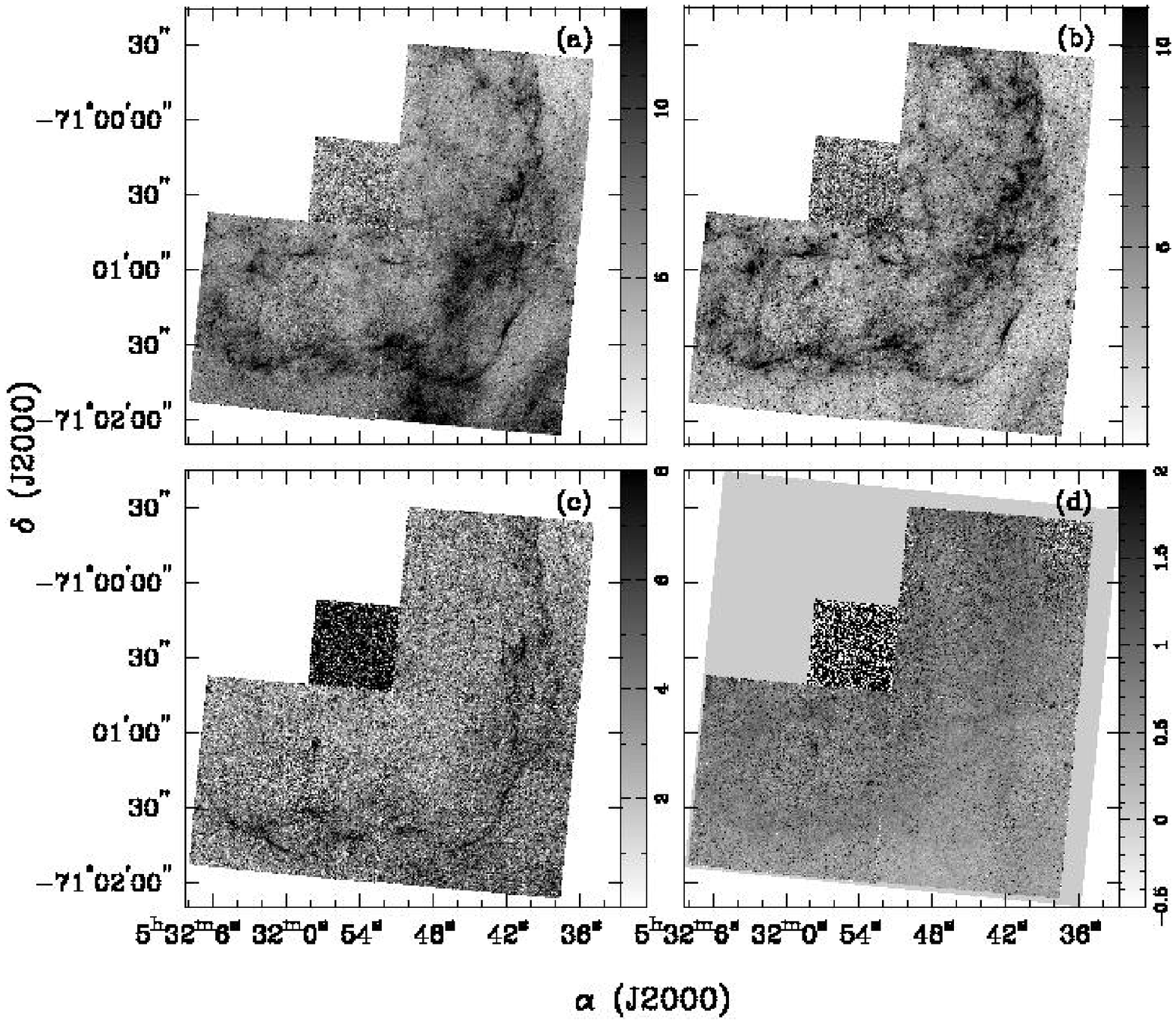}
\caption{HST WFPC2 mosaicked images of the southwest side of N206
in (a) \ha, (b) [\ion{S}{2}], (c) [\ion{O}{3}] and 
(d) [\ion{S}{2}]/\ha\ ratio map.}
\label{fig:hst3band}
\end{figure}

The [\ion{O}{3}] emission, as seen in Figure~\ref{fig:hst3band}c, tends 
to the outer edges of the filaments.  [\ion{O}{3}] is a useful ``tracer" 
of shock fronts, due to the relatively narrow range of temperatures and 
densities in which this emission dominates.  Thus the tendency of the 
[\ion{O}{3}] to lie in the limbward direction of the filaments implies a 
fairly regular expansion. There do not appear to be any  counterparts 
in \ha\ or [\ion{O}{3}] to the linear feature seen in radio by \citet{K+02}.  
The \ha\ and [\ion{S}{2}] morphologies, shown in Figure~\ref{fig:hst3band}a
and b,  are quite similar, including the filamentary structure, as is 
typical for cooled ($\sim$10$^4$ K) post-shock gas. 

\subsection{Hot Gas in the SNR's Interior}

\begin{figure}
\epsscale{1.0}
\plotone{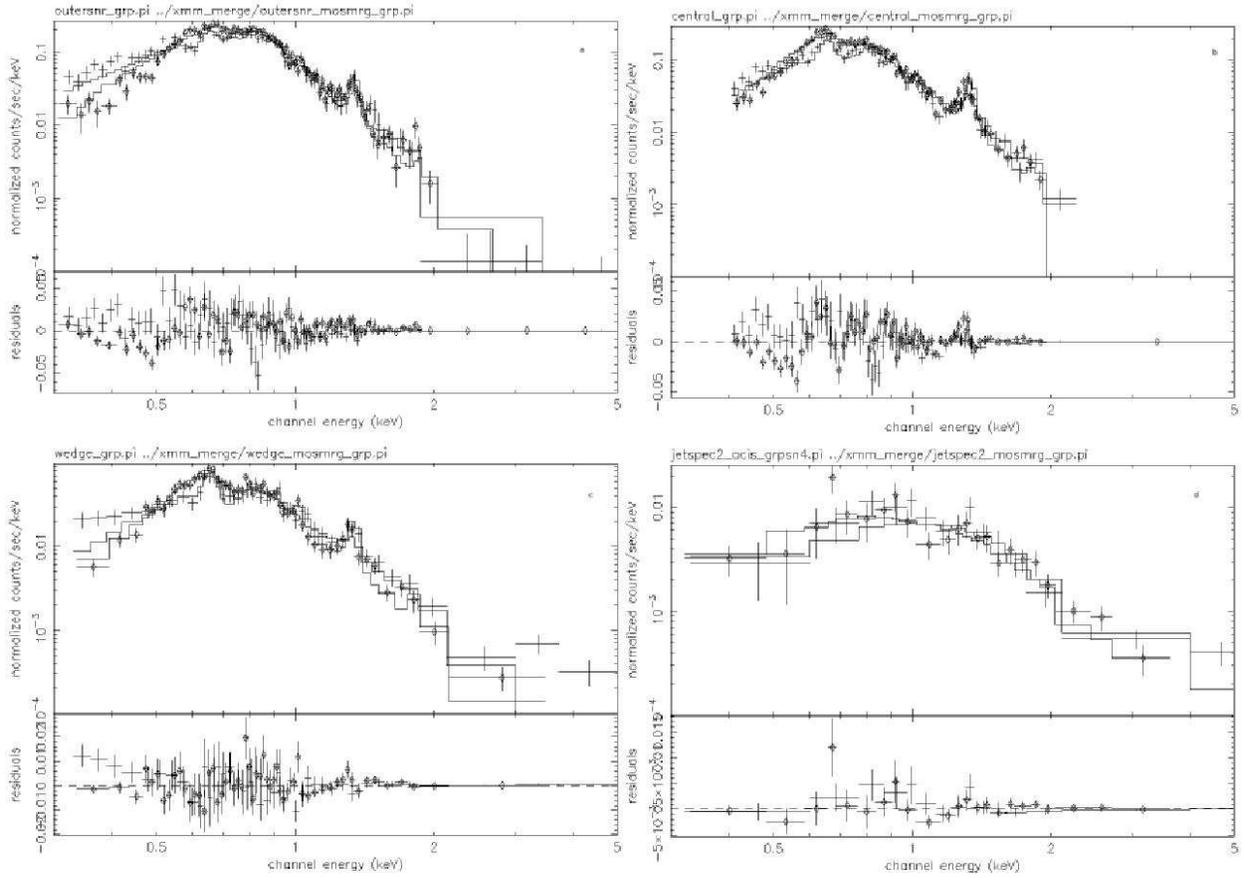}
\caption{Chandra ACIS (crosses) and XMM-Newton EPIC-MOS (dotted crosses) spectra, fits and residuals for four extraction regions: (a) Region 2, outer limb; (b) Region 3, central; (c) Region 4, wedge; (d) Region 5, linear feature.}
\label{fig:xray_spec}
\end{figure}

\begin{figure}
\epsscale{1.0}
\plotone{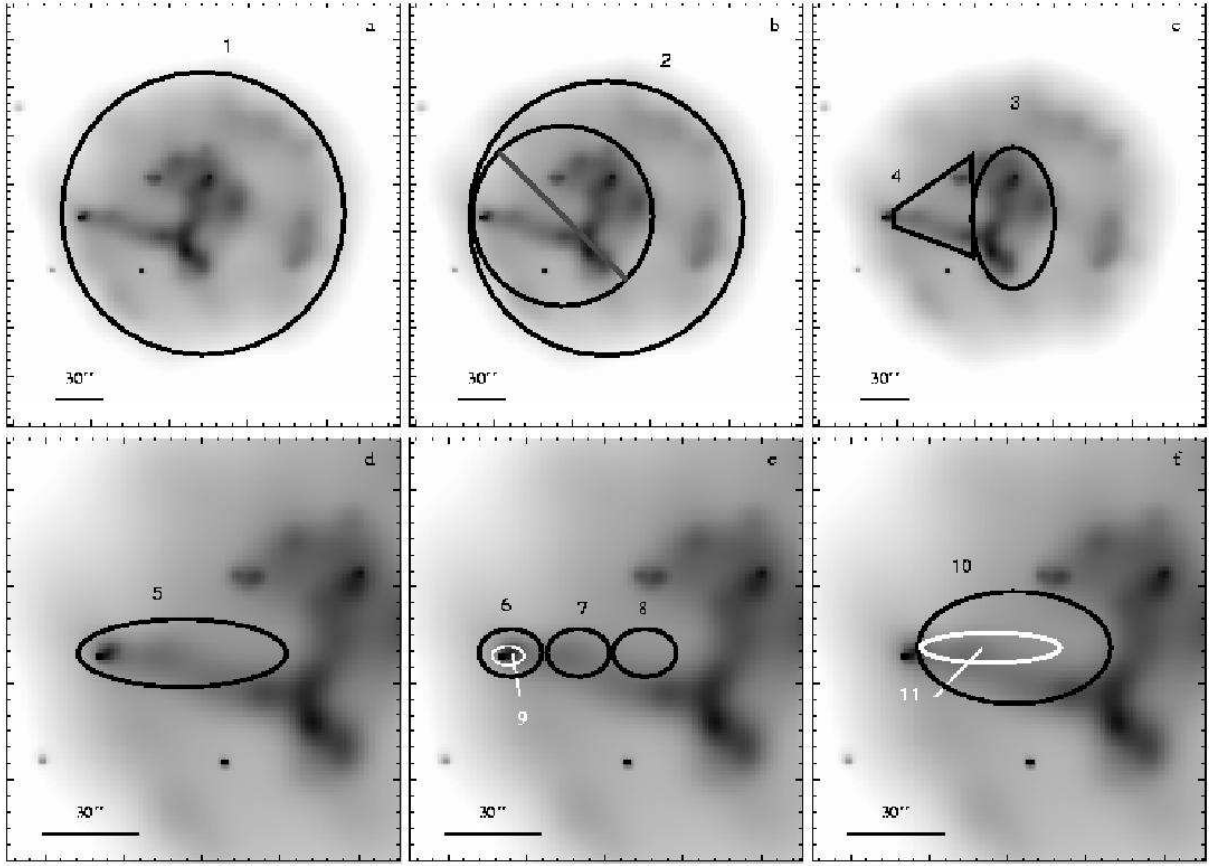}
\caption{Smoothed Chandra ACIS image with regions for spectral analysis, as 
listed in Table~\ref{tab:reglist}, overlaid and labeled.  (a) Region 1; 
(b) Region 2; (c) Regions 3-4; (d) Region 5;  (e) Regions 6-9; (f) Regions 10-11.}
\label{fig:xray_regions}
\end{figure}

Spectral fits to the X-ray data from the various instruments (\chandra\ 
ACIS, \xmm\ EPIC-MOS and EPIC-pn) show that the emission from the SNR is 
dominated by thermal emission; a number of thermal emission lines are 
visible (Fig.~\ref{fig:xray_spec}). We therefore utilized thermal plasma 
models to determine the plasma parameters. As it is uncertain whether the 
plasma has reached ionization equilibrium, we fit both a non-equilibrium 
ionization model (NEI; ``pshock" in \textsc{xspec}) and a collisional-ionization 
equilibrium model (CIE; ``MeKaL" in \textsc{xspec})\footnote{Details and 
references for the pshock and MeKaL models can be found at \\ 
\url{http://heasarc.gsfc.nasa.gov/docs/xanadu/xspec/manual/node39.html
and node40.html.}}. We selected the following regions for further analysis:
(Region 1) one comprising the entire SNR for comparison with other X-ray studies; 
(2) a region enclosing much of the SNR limb, which is expected to be dominated
by emission from recently shocked gas; (3) a region enclosing the central X-ray
brightening but excluding emission thought to be associated with the ``linear
feature"; (4) a region covering an apparent wedge-shaped X-ray brightening north 
and south of the ``linear feature", excluding a possible point source; (5) a 
region corresponding to the ``linear feature" seen in radio; (6-8) three 
equal-area regions along this feature to examine possible changes over its 
length; (9) a small region surrounding a possible point source at one end of 
the ``linear feature"; (10) a region covering emission immediately behind the 
possible point source, excluding a narrow area corresponding to the brightest
radio emission from the ``linear feature"; and (11) a region covering the 
aforementioned radio-bright linear feature. A list of the regions used for 
X-ray analysis is given in Table~\ref{tab:reglist} and shown in
Figure~\ref{fig:xray_regions}.

We chose two of these regions thought to include only thermal X-ray emission:  
``Outer Limb" (Region 2, Fig.~\ref{fig:xray_regions}b) and ``Central" (Region 
3, Fig.~\ref{fig:xray_regions}c). We compared the fits of simple thermal models 
to the data from each of the X-ray instruments used (Table~\ref{tab:snrspecfit}).  
Best fits are obtained with sub-solar metal abundances, consistent with the 
ambient metal abundances in the LMC of about 30\% solar \citep{RD92}. Fits to 
the absorption column density are reasonably consistent for regions with 
good statistics, giving $N_{\rm H}$ = 3$\pm$1 $\times$ 10$^{21}$ cm$^{-2}$.  
Fits to the temperature differ by a factor of two depending on whether CIE or 
NEI is assumed; CIE fits have $kT$ = 0.23$\pm$0.01 keV and NEI fits have  
$kT$ = 0.4$\pm$0.1 keV (Table~\ref{tab:snrspecfit}). The fits to the 
ionization parameter $\tau$ ($=nt$, where $n$ is the hot gas density and 
$t$ is the time since ionization) indicate that the hot gas deviates 
considerably from ionization equilibrium, with 
$\tau$ = 3$\pm1 \times 10^{11}$ cm$^{-3}$ s.  In general, the NEI 
model also provides a better statistical fit to the data. We will 
therefore refer primarily to the results of the NEI fits in the 
subsequent discussion.

\subsubsection{Overall SNR Characteristics}

In order to facilitate comparison of our X-ray results with other X-ray 
studies at lower resolution, we include the ``Whole SNR" region
(Region 1 in Table~\ref{tab:reglist}, Fig.~\ref{fig:xray_regions}a), which 
includes all X-ray emission from the remnant, including that from the 
area of the ``linear feature".  These fits (Table~\ref{tab:wholesnrspec}) 
largely fall within the range described above for the ``Outer Limb" and 
``Central" regions, indicating that the SNR's emission is dominated by 
other sources of emission than that of the ``linear feature."  While an 
additional power-law component with a normalization of one-third that 
of the thermal plasma component slightly improves the fit, this 
improvement is not significant.  The NEI fit yields an absorbed flux of 
7$\pm 2 \times 10^{-13}$ erg cm$^{-2}$ s$^{-1}$, an unabsorbed flux of 
4$\pm 2 \times 10^{-12}$ erg cm$^{-2}$ s$^{-1}$, and a luminosity of 
8$\pm 4 \times 10^{35}$ erg s$^{-1}$, all over the 0.3$-$8.0 keV range.

The normalization of the NEI model fits to X-ray data can be used to 
calculate the rms electron density ($n_{\rm e}$) of the hot gas within the SNR.  
We assume the remnant to be roughly spherical, with a radius of 21 pc, 
at a distance of 50 kpc.  We further assume that the gas is composed of 
fully ionized hydrogen and helium ($n_{\rm e} \sim 1.2 n_{\rm H}$).  Then, using 
the normalization from NEI fits to the Chandra ACIS data for the ``Whole SNR"  
region (Table~\ref{tab:wholesnrspec}) we find a density in the hot gas of 
$n_e=0.24\pm0.05$ $f_{\rm hot}^{-\half}$ cm$^{-3}$.  Using the ``Outer" and 
``Central" regions (Table~\ref{tab:outerlimbspec}-\ref{tab:centerspec}), 
which do not include emission from the area around the ``linear feature," 
for this calculation gives a very similar density of $n_{\rm e}=0.23\pm0.09$ 
$f_{\rm hot}^{-\half}$ cm$^{-3}$.  The parameter $f_{\rm hot}$ is a volume 
filling factor for the hot gas. If we assume $f_{\rm hot}$=1, to reflect the 
centrally filled X-ray morphology of the remnant, the hot gas density is 
$n_{\rm e}=0.24\pm0.09$ cm$^{-3}$; if the gas only partially fills this volume, 
the density will rise in inverse proportion to the square root of this 
filling factor.

From this derived density of the X-ray emitting material, we obtain a
mass of 5.3$\pm 0.2\times 10^{35}$ $f_{\rm hot}^{\half}$ g, or 
270$\pm$10 $f_{\rm hot}^{\half}$ M$_{\sun}$.  Presuming that the hot gas 
fills the entire remnant ($f_{\rm hot}$=1), this mass is greater than that 
expected from SN ejecta alone, emphasizing that N206 is an older SNR 
whose emission is dominated by swept-up material from the surroundings; 
and supports the LMC-like abundances found from the spectral fit.  Using 
the temperature and density from spectral fits to the X-ray data, we find 
a thermal pressure, $P_{\rm th} = nkT$, of 3.3$\pm 0.4 \times 10^{-10}$ 
$f_{\rm hot}^{-\half}$ dyne cm$^{-2}$ in the hot gas. As the analysis 
in \S 3.2 will show, this is significantly greater than the thermal 
pressure we find for the cool shell.  We can also calculate the thermal 
energy in this gas, $E_{\rm th} = 1.5 nfVkT$.  This gives a thermal energy 
of 6$\pm 1 \times 10^{50}$ $f_{\rm hot}^{\half}$ erg for the hot gas. 
Finally, we can estimate the shock velocity under the (somewhat dubious) 
assumption that the bulk of the X-ray emission is being generated at the
current shock front, using $kT = (3/16) \mu v_{shock}^2$, with the reduced 
mass $\mu= 0.61$ (for He:H = 1:10).  If we presume the newly shocked gas 
to be at 0.45 keV, we find that such a temperature would result from a 
shock speed of 620 km s$^{-1}$, implying an overall expansion velocity of 
470 km s$^{-1}$. For simplicity, we have assumed in the above calculations 
that the ions have equilibrated with the electrons; although this assumption 
may well not be valid, it provides us with at least initial estimates.  The 
derived energy and pressure are similar to those for other SNRs in the 
adiabatic stage of evolution.  

\subsubsection{Specific Hot Gas Features}

N206 is clearly more complex than the simple shell approximated above.
It features an increase in surface brightness toward the remnant center,
a ``wedge" of emission appearing to surround the ``linear feature", and 
the feature itself, terminating in a possible compact source.
The properties of these features will be discussed in more detail in \S4.

The central brightening on the SNR is quite puzzling. The spectrum 
of this region (Table~\ref{tab:centerspec}) appears similar to that 
for the limb of the remnant, with slightly higher abundances, particularly
oxygen.  There is no evidence for a strong power-law component in this 
region; fits with such a component require it to be normalized to a 
small fraction of the thermal component's emissivity.

Toward the eastern side of the remnant, surrounding the ``linear 
feature," the X-rays show a slightly brighter broad ``wedge" of 
emission (Region 4 in Table~\ref{tab:reglist}, Fig.~\ref{fig:xray_regions}c).  
Fits to the spectrum of this region (Table~\ref{tab:wedgespec};
this spectrum does not include contributions near the ``compact source")
suggest a markedly higher plasma temperature than elsewhere in the 
SNR.  While a nonthermal contribution might artificially skew a thermal 
model fit toward higher temperatures, fits which included nonthermal
emission over a tenth of the thermal emissivity could be statistically
ruled out at the 90\% level.

We also examined a region covering the area of the ``linear feature", as
seen from radio images (Region 5 in Table~\ref{tab:reglist}, Fig.~\ref{fig:xray_regions}d).  We defined a hardness ratio (H$-$S/H+S) 
such that S=0.3-1.0 keV and H=1.0-8.0 keV (Table~\ref{tab:reglist}).  Using 
these hardness ratios to compare this region to others throughout the SNR,
we find that, while the contribution of soft-X-rays still dominates, the
proportion of hard X-rays in this region is greater than elsewhere in the
SNR.  This increase in the hardness ratio is consistent with the combination 
of thermal emission from the SNR and emission from a harder source, as
for example the small-diameter bright source at the tip of this region.
While a localized increase in temperature could also explain this hardening,
it would be quite coincidental that this increase appears only in this 
portion of the SNR that also shows the small-diameter X-ray source and 
elongated radio feature (Fig.~\ref{fig:xmm_acis}c-e).

We performed spectral fits to data from Region 5, though we note that these 
fits are somewhat limited by the relatively low number of counts. We find that 
either a thermal plasma model or a power-law model can provide a statistically 
adequate fit to this source (Table~\ref{tab:linearspec}). The ``best" fit  
arises from a combination of nonequilibrium thermal plasma and power-law 
models, providing roughly equal contributions to the spectrum.  Although the 
improvement in the fit is not statistically significant, such a combination is 
consistent with the scenario inferred from the hardness ratios.   To examine the 
detailed properties of the ``linear feature", we used the \chandra\ ACIS data 
to analyze emission from smaller segments along that feature (Regions 6-8 in 
Table~\ref{tab:reglist}, Fig.~\ref{fig:xray_regions}e).  These fits, summarized 
in Table~\ref{tab:smregspec}, should be considered as preliminary estimates only.

\subsection{The Cool Shell of the SNR}

Figure~\ref{fig:hst3band}d shows an [\ion{S}{2}]/\ha\ ratio map of the
southwestern side of N206. [\ion{S}{2}]/\ha\ ratios range from 0.7 to 1.2 
across the SNR, typical for shocked gas, with an average value of 0.9.  
This is initially somewhat surprising, as it indicates that there has not 
been much ionization of material from the nearby OB association. However, 
\hi\ maps of the vicinity show an \hi\ shell, with the N206 \hii\ region
largely situated within the central cavity, so that the \hi\ shell walls 
probably block most of the ionizing radiation before it can reach the 
SNR \citep{DCS05}. 

We measured the average \ha\ surface brightness of filaments in the 
flux-calibrated WFPC2 image of N206 to be  (0.8-1.8)$\times$10$^{-17}$ 
erg cm$^{-2}$ s$^{-1}$ pix$^{-1}$, or (0.8-1.8)$\times 10^{-15}$ erg 
s$^{-1}$ cm$^{-2}$ arcsec$^{-2}$. Assuming a cool shell temperature 
of 10$^4$ K, this surface brightness implies an optical emission measure 
of 680$\pm$ 260 pc cm$^{-6}$.  We presumed the average filament thickness 
along the line of sight (20\arcsec\ $-$ 30\arcsec) to be equal to its 
width perpendicular to the line of sight, and used this as a 
representative number for the path length ${\cal L}$ through the warm 
ionized gas.  Using this ${\cal L}$, we calculate an rms electron 
density in the shell of about 10$\pm$4 cm$^{-3}$.   If we presume the 
SNR to be in the adiabatic ``point-blast" stage of expansion \citep{S59}, 
and the cool shell to be representative of the ambient ISM, we would 
expect the ambient ISM density to be roughly one-quarter of that in 
the shell, or about 3$\pm$1 cm$^{-3}$.   

\begin{figure}
\epsscale{1.0}
\plotone{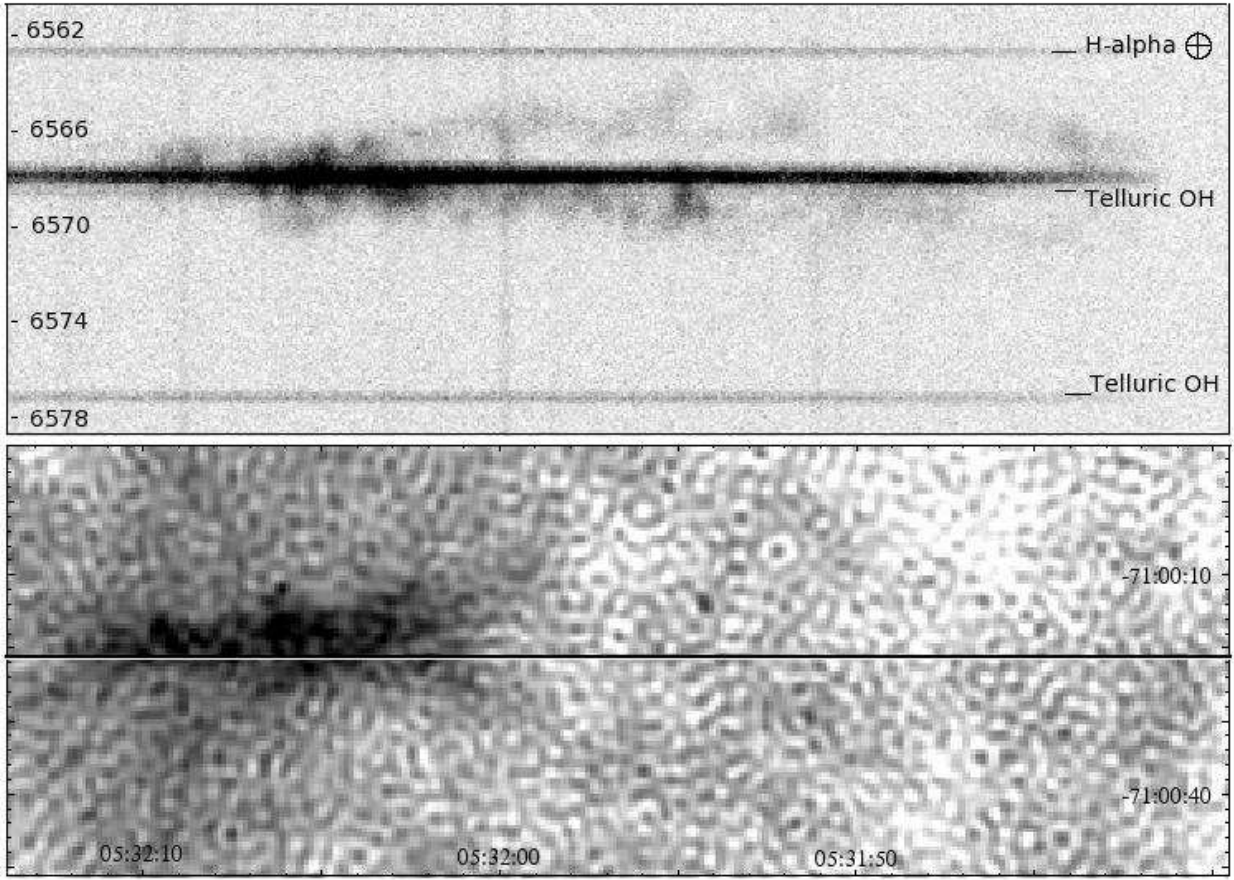}
\caption{Top: echelle spectrum  along ``linear feature". Bottom:
ATCA 6 cm radio image of ``linear feature" showing slit location.}
\label{fig:echelle}
\end{figure}

The optical echelle data (Fig.~\ref{fig:echelle}) show several lines in 
the \ha\ spectral region, including the narrow geocoronal \ha\ (6562.85 \AA) 
and telluric OH 6-1 P2(3.5) 6568.779 and 6-1 P1(4.5)e/f 6577.183/.386 
lines \citep{O+96}, as well as broader lines corresponding to 
Doppler-shifted nebular \ha\ emission toward the SNR.  The latter 
include both a velocity component constant along the slit, showing the 
systemic velocity of the SNR, and the characteristic bow-shaped pattern 
deviating from this systemic velocity, showing motions within the 
expanding gas.

In order to measure the systemic velocity, we extracted a velocity profile
from a region 13\arcsec\ wide outside of the emission of the SNR expansion
pattern. The profile showed two components, one of which was identified as 
the telluric OH line.  The other component is the H$\alpha$ line with a 
measured Doppler shift of 241$\pm$4 km s$^{-1}$; as the SNR expansion 
pattern appears to converge to this component, we take this as the systemic 
velocity for the N206 SNR. For comparison, we note that the nearest position 
to N206 in the \hi\ maps of \citet{R+84} shows velocity components at 
240$\pm$7 and 262$\pm$26 km s$^{-1}$.  

To characterize the expansion of the N206 SNR, we measure the 
velocity offsets ($\Delta v$) from the systemic velocity in both the blue
and red directions.  As the material at the forefront of the SNR expansion
may be reasonably expected to show the highest velocity, we take the
greatest of these velocity offsets to represent the overall expansion
velocity of the remnant.  The greatest offset in the blue direction 
($\Delta v_{\rm blue}$) is $-$202$\pm$5 km s$^{-1}$, while the largest 
$\Delta v_{\rm red}$= $+$193$\pm$5 km s$^{-1}$. We therefore estimate the
expansion velocity $v_{\rm exp}$ = 202$\pm$5 km s$^{-1}$. Using this value 
for expansion, and assuming that the SNR is in the Sedov phase, we would 
expect the shock velocity 
$v_{\rm shock}$ = $4/3\ v_{\rm exp}$ = 270$\pm$5 km s$^{-1}$.
It should be noted that the errors given here are only the random errors 
of the measurements.   

It is possible that some of the optically emitting material is too faint 
for detection in these observations. If the highest-velocity material is 
undetected, the actual expansion velocity may be higher than our estimate. 
For example,  \citet{CK88} found a higher expansion rate of 250 km s$^{-1}$; 
the discrepancy may be due to the difficulty of discerning motions in the 
faint outer material.   To an extent, this may also apply to the large 
discrepancy between the shock velocity calculated from the temperature of 
the hot gas in \S 3.1.1, and that measured from the warm ionized gas.  More 
likely, however, this discrepancy is due to two factors. (1) The X-ray 
temperature may not be a reliable representation of the newly shocked gas, 
particularly given its large deviation from a limb-brightened shell morphology. 
(2) We expect the X-ray emission to arise from areas where the shock front is 
moving through diffuse gas, while the optical emission is expected to be 
generated where the shock is moving through higher density gas, as seen in 
the relative densities of the hot gas to the warm ionized gas.  Thus, we 
expect that the shock within the optically-emitting clumps has been 
somewhat slowed by its progress through this denser material.  The overall
expansion velocity from the blast wave is probably intermediate between the
measured optical expansion of 202 km s$^{-1}$ and the calculated expansion
from X-ray temperatures of 470 km s$^{-1}$.

Using the shell electron density of n=10 cm$^{-3}$ calculated above, 
and presuming $n_{\rm He}/n_{\rm H} = 0.1$ and singly ionized He 
($n_e = 1.1 n_{\rm H}$), we can infer the mass of gas in the shell 
according to $M = n m_{\rm H} V_{\rm shell}$, where $V_{\rm shell}$ 
is the shell volume.  
From the filamentary structure at the edge of the shell, we measure an 
average shell thickness of about 1\farcs1$\pm$0\farcs5 (0.27$\pm$0.13 pc 
at a distance of $\sim$50 kpc to the LMC). If we assume a simple spherical 
shell of this thickness, we calculate a shell mass of 
9$\pm 6 \times 10^{35}$ g, or 460$\pm$300 M$_{\sun}$. This is almost 
certainly an underestimate, as the filamentary structure shows a much 
greater extent of cool gas than such a simplistic scenario. If instead 
we presume the cool gas occupies a volume filling factor of as much as 
0.1, we obtain a mass estimate of 2.4$\pm 1.7 \times 10^{36}$ g, or 
1200$\pm$800 M$_{\sun}$.  

Using these as low- and high-end estimates of the mass range, and the 
expansion velocity above, we calculate the kinetic energy in the cool
shell, $0.5 M_{\rm shell} v_{\rm exp}^2$, to find values of 2$\pm 1\times 10^{50}$
erg and 5$\pm 3\times 10^{50}$ erg, respectively.  One can also use the 
density calculated above to calculate the thermal pressure in the shell,
$P = nkT$, where $k$ is the Boltzmann constant and $T$ is the temperature,
presumed here to be $10^4$ K.  This equation gives a pressure of 
3$\pm 1\times 10^{-11}$ dyne cm$^{-2}$.  Again, these values for kinetic 
energy and shell pressure are typical for middle-aged SNRs.
The physical characteristics of the N206 SNR are summarized in 
Table~\ref{tab:parameters}.

\section{Specific Radio and X-ray Features}

The high spatial resolution of \chandra's ACIS allows us to
examine specific regions in more detail.  We have isolated three features
that appear to be of particular interest: X-ray emission associated with
the radio-identified ``linear feature;" the X-ray knot and
surrounding emission which coincides with the very tip of that linear
feature; and the region of enhanced X-ray surface brightness toward the
center of the SNR.  Below, we describe each of these features individually
and then discuss the possible origins for these features.

\subsection{Central Emission}
 
The interior X-ray emission is brightest from an irregular region 
extending N$-$S for half the diameter of the remnant.
This central emission is clearly dominated by
thermal X-rays, with prominent line features including the blends of 
He-like lines of Mg and Si.  The emission is largely soft ($<$ 2 keV) 
and thermal plasma fits to data from this region give parameters for 
$N_{\rm H}$ and $kT$ similar to those found for the SNR as a whole.  
However, there does appear to be some difference in abundance 
distributions between the central area and the outer regions of the
SNR (excluding emission associated with the ``linear feature"). The 
oxygen lines in the central region are more prominent, with respect 
to the iron blends (Fe L), than in other regions of the SNR 
(Table~\ref{tab:centerspec}).  These differences should be treated 
with caution, as there are large uncertainties in the abundance
determinations; however, the difference in oxygen abundance between the
two regions within the SNR is significant when compared to the 90\% error
ranges for this quantity.  If the central emission is due to ``fossil
radiation," i.e. gas that was shock-heated during earlier phases of the
SNR expansion, we would expect that this gas
would have a higher proportion of ejecta to swept-up matter than the more
recently-shocked gas, and therefore that the higher oxygen abundance 
reflects the presence of oxygen-rich ejecta, as seen in the much younger
mixed-morphology SNR 0103-72 in the SMC \citep{P+03}.  The presence of 
O-rich material cannot be confirmed from \hst\ [\ion{O}{3}] maps. This
probably reflects the fact that the X-ray emission which shows the oxygen
excess is primarily from the hot central cavity, where the temperatures
are too high and the densities too low for [\ion{O}{3}] to show up.

It is worth noting that in addition to the similar oxygen-rich and 
mixed-morphology SNR 0103-72 \citep{P+03}, the oxygen-rich SNR N132D 
shows, likewise, a higher ratio of oxygen (O$^{7+}$) to other elements 
in the remnant interior than at the bright rim \citep{B+01}.  
Nucleosynthesis models indicate that a high oxygen abundance within a
SNR's ejecta is the result of a Type II SN \citep[e.g.,][]{T+95}.
The examples of ``mixed-morphology" SNRs which show signs of enhanced
oxygen abundances, therefore, may imply that the ``mixed-morphology" 
SNRs are likely to have originated from Type II SNe.

As with other mixed-morphology SNRs \citep{RP98},
the expanding shock of N206 has slowed to the point where bright X-rays 
from the limb do not dominate the overall emission from the remnant.
Thus, it is reasonable to think that N206 follows the typical pattern
for mixed-morphology SNRs, wherein the central emission is dominated 
by fossil radiation from earlier large-scale shock heating during the 
SNR's evolution.

\subsection{Linear Feature}

An elliptical region (65\arcsec\ E$-$W $\times$ 21\arcsec\ N$-$S) 
surrounding the ``linear feature" in radio was selected from the 6 cm 
radio image \citep{K+02}, and was used to extract spectra from the 
corresponding X-ray data (Region 5).  Comparison of the hardness ratio 
for this region to those elsewhere in the SNR, as discussed in \S3.1.2, 
suggests the presence of a hard component in addition to the soft 
emission seen throughout N206.   A combined NEI and power-law model with 
$kT=0.4 \pm 0.2$ and $\Gamma = 2.2 \pm 0.2$ provided the best fit. 
Fits to these spectra are summarized in Table~\ref{tab:linearspec}.  
(Note that the value of $\tau$ obtained for this fit would indicate
a plasma in collisional ionization equilibrium; we retain the NEI 
model, however, for consistency in intercomparison with the other fits.)
Using these fitted values, we obtained an absorbed X-ray flux of 
4$\pm 1 \times$ 10$^{-14}$ erg cm$^{-2}$ s$^{-1}$ over the 0.3-8.0 keV 
range, which corresponds to an unabsorbed flux of 1$\times$ 10$^{-13}$ 
erg cm$^{-2}$ s$^{-1}$. At the LMC distance of 50 kpc, this gives a 
luminosity of 3$\times$ 10$^{34}$ erg s$^{-1}$ over this energy range.

To further examine the nature of this linear feature, a series of three
equal-area regions (20\arcsec $\times$ 15\arcsec) E$-$W along the feature 
were identified in \chandra\ images for further spectral examination (Regions
6-8 in Table~\ref{tab:reglist}). Comparison of the hardness ratios defined 
above for regions 6-8 shows a pronounced shift from regions dominated by 
high-energy photons to those dominated by low-energy photons. 
 
We also performed spectral fits for these regions (Table~\ref{tab:smregspec}).  
It should be stressed that, due to the low numbers of counts, these spectral 
fits are quite uncertain.  Cited values of $\chi^2_{\rm red}$ are low due to 
the large error bars for the spectral bins; however, further binning was 
deemed undesirable due to the loss of remaining spectral information.  However, 
as with the hardness ratios discussed above, the fits illustrate the shift from 
harder to softer emission along the feature.  

In order to see whether the relative fluxes of thermal to nonthermal 
emission would change substantially as one moved from the remnant center 
toward the X-ray bright knot at the east end of the linear feature, we
adopted a dual approach to these spectral fits. Initially, we fit a 
two-component (power-law and NEI) model jointly to all three regions, 
and determined the flux in each component for each region.  As an 
alternate approach, we fit the power-law component to the high-energy 
end of the spectrum only ($>$2 keV) and fixed those parameters, then 
fit the joint model to the low-energy end of the spectrum. Again, the 
flux in each component was determined.  The results, summarized in 
Table~\ref{tab:linearflux}, indicate that the X-ray emission along 
the linear feature is more strongly dominated by nonthermal emission 
the further one moves eastward (closer to the knot) along the feature, 
and dies off on the west side.  While this result is largely qualitative, 
given the errors, we argue that the increase in ``hard" emission toward 
the knot is most likely to result from a nonthermal source.  

The nonthermal spectra of the X-ray emission and radio emission in the 
``linear feature" imply the presence of a pulsar-wind nebula (PWN).  
The most likely source for nonthermal X-rays within a SNR is synchrotron 
radiation. While nonthermal X-ray emission may occasionally be generated 
by a fast SNR shock at the outer limb of a SNR, both the relatively slow 
expansion of this SNR and the shape of the emitting region of the 
nonthermal emission, i.e. over a small spatial segment perpendicular to 
the shell, argue against this scenario. Likewise, it seems unlikely that
the motion of a compact source would be sufficient to produce a bow 
shock capable of generating significant synchrotron X-rays.  The most 
plausible explanation, then, is  that the acceleration of particles by a 
``hidden" pulsar is the source of the nonthermal emission (in both radio 
and X-rays) seen in the linear feature.

The westward decrease in the ratio of nonthermal to thermal X-ray flux 
along the linear feature suggests that it is a PWN generated by an eastward  
moving pulsar. The lifetime of the high-energy particles generated 
in a PWN is comparatively short; the radio-emitting particles persist
and so trail farther behind the X-ray extent.  The radio pulsar 
observations were unable to find any periodicity greater than 2 ms 
to the radio signal from the area of the ``linear feature".  It is by 
no means uncommon for a PWN to be detected in radio and X-rays without 
a radio point-source counterpart; for example, N157B, also in the LMC, 
was deduced to have a PWN long before the X-ray pulsar was discovered, 
and a radio counterpart for that pulsar has yet to be found \citep{W+01}.

Another possibility that must be considered, of course, is that these  
features are associated with a background source rather than the SNR.
\citet{K+02} discuss this scenario, but conclude that this is unlikely.  
The presence of extended nonthermal X-ray emission provides an additional 
reason to believe the feature is not a background source, as such a 
feature would be highly unusual in a background galaxy.

If we accept the premise that the ``linear feature" is a PWN, we may 
anticipate that in part, the morphology of the PWN is created by a 
bow-shock structure.  The dynamics of a rapidly moving pulsar and 
evolving SNR lead to the prediction of the formation of a bow shock 
when the SNR expansion speed has decelerated sufficiently for the pulsar's 
motion relative to the shocked SNR material to become supersonic 
\citep[e.g.][]{V+98}. At this point the PWN is deformed, leading to a
substantial offset between the pulsar and the center of the PWN; this
is certainly consistent with the observed bright knot at the 
outer tip of the ``linear feature" presumptive PWN.

Numerical simulations predict that for a remnant in the Sedov stage, 
this should occur when the distance traveled by the pulsar from the center
of the SNR, $R_{\rm PSR}$, approaches the SNR radius $R_{\rm SNR}$ 
according to $R_{\rm PSR}$/$R_{\rm SNR}$ $\ga$ 0.677 \citep{V04,V+98}.
Presuming that the geometric center of the SNR shell accurately 
represents the site of the SN, the current position of the small-diameter
X-ray source is such that the transverse component of $R_{\rm PSR}$ is
0.85 $R_{\rm SNR}$, well over the point for bow shock formation. If 
the direction of pulsar motion is not perpendicular to the line of 
sight, $R_{\rm PSR}$ increases accordingly.  These findings are in 
general accord with those of \citet{K+02}, and their assumptions in 
their estimation of the pulsar velocity and age.

The fact that emission from the ``linear feature" in X-ray and radio does
not dominate the overall emission from the remnant presents an additional
question.  Its spectral index would place it among the most flat radio 
SNR shells, although overlapping (within the error bars) with remnants
such as IC443 \citep{Kawa+02}. It is difficult to see why the SNR as a 
whole should show a radio index typical of filled-center remnants 
\citep{K+02} if the putative PWN is largely confined to this feature.  
However, the radio emission from N206 is faint compared to that from other 
SNRs, and therefore the uncertainties in the spectral index determination 
are considerable. In addition, residual emission from the N206 \hii\ region
may contaminate the radio emission, adding additional uncertainty.

\subsection{Evidence for a Compact Source}

In the \chandra\ images, a distinct knot of X-ray emission appears 
to the far eastward end of the linear feature, near the SNR limb. To 
investigate whether this knot is consistent with an unresolved source, 
we generated a point source of comparable X-ray flux using the MARX 
simulator, and compared the profile of this simulated source to that 
of the knot.  We conclude that the knot is a small diffuse region of
dimension 2\arcsec. This can also by seen by comparison with the 
appearance of the moderately bright point source $\sim$50\arcsec\ SW
of the knot (identified above with 2E 0532.6-7102). If a point source
is embedded within the knot, only $\le$50\% of the counts from 
the knot could be due to that point source.

We extracted counts within a 2\arcsec\ radius of the knot from the 
merged \chandra\ ACIS events file (Region 9 in Table~\ref{tab:reglist}, Fig.~\ref{fig:xray_regions}e),  obtaining a count rate after
background subtraction of 7.9 $\times$ 10$^{-4}$ ct s$^{-1}$, or 
52 counts over the 65.6 ks exposure time, an insufficient quantity 
to obtain a meaningful spectrum. Fig.~\ref{fig:xmm_acis}, however, 
shows that the knot shows noticeably harder emission than that from 
the surrounding SNR. Once again using the hardness ratio defined above,  (Table~\ref{tab:reglist}), we find a hardness ratio of 0.5 for the 
knot, in contrast to hardness ratios of -0.45 and -0.58 for the ``outer" 
(Region 2; SNR limb excluding the putative PWN) and ``central" (Region 3) 
areas.

To obtain a first-order estimate of spectral properties for this knot, 
we fixed $N_{\rm H}$ (and abundances, for the thermal case) to the 
values determined by fits to the X-ray emission from the rest of the 
SNR.  The data are consistent with a thermal plasma model with 
temperatures $>$ 1 keV, or a power-law model  with a spectral index 
$\Gamma$ between 2 and 3.  Using a power-law model with $\Gamma=2.4$, 
we obtain an estimate for the absorbed flux of 6$\times$ 10$^{-15}$ 
erg cm$^{-2}$ s$^{-1}$, an unabsorbed flux of 1$\times$ 10$^{-14}$ 
erg cm$^{-2}$ s$^{-1}$, and a luminosity of 3$\times$ 10$^{33}$ 
erg s$^{-1}$ at the LMC distance, all over the 0.3-8.0 keV energy 
range.  The maximum luminosity for an embedded point source is 
therefore $\sim$1.5$\times$ 10$^{33}$ erg s$^{-1}$ over this range.

An elliptical region (5\arcsec$\times$8\arcsec) immediately west of 
the knot, along the linear feature, yielded a count rate of 
1.4 $\times$ 10$^{-3}$ ct s$^{-1}$. This region has a very similar 
spectrum to that from the knot, suggesting that the spectrum 
determined from the knot is dominated by emission from its 
immediate surroundings. Combining the spectra from these two regions 
allowed us to increase the signal-to-noise ratio for a slightly better 
spectral fit.  While these fits would still not allow us to rule out a 
thermal plasma interpretation at 90\% confidence, they confirm that a 
relatively high temperature $kT > 1$ keV, is required. More 
plausible is a power-law fit, with $\Gamma = 2 \pm 1$, roughly 
consistent with that measured from the ``linear feature" as a whole.
While fits to smaller regions along the linear feature suggest a
possible spatial variation in the power-law spectral index, the
sensitivity of the X-ray observations is insufficient to make this
determination at a statistically significant level.

A brief analysis of the X-ray power spectrum shows no evidence for 
periodic emission from an embedded pulsar. However, the timing 
resolution of these observations, 3.2 seconds, would not be sufficient
to detect the pulsations from a typical pulsar. In addition, the 
estimated maximum of 25 counts from a point source is insufficient 
to determine a period for its emission, even if the timing resolution 
were available.

\subsection{Possible Bow Shock}

Proceeding westward from the knot, X-ray emission associated with 
the ``linear feature" broadens into a wedge of brighter X-ray emission 
which then appears to merge with the central emission.   Faint 
emission connects this patch to the compact knot close to the eastern 
rim.  We interpret this connection as a ``bow shock," clear in the south
and barely discernable in the north. This bow shock merges with the
radio ``linear feature" close to the knot. A bow shock in front of a 
hypothetical moving pulsar might re-heat the material through 
which it moves, leaving a trail of ``fossil radiation" back to the 
remnant center and merging there with fossil radiation from much 
earlier fast-moving shocks.  

One approach to these features would be to consider much of the X-ray 
emission between the bright linear feature and the central region as 
the actual boundary of the bow-shocked gas. In this case we would have 
a much larger opening angle for this structure than that found by 
\citet{K+02}. Using the X-ray opening angle of about 
57\degr $\pm$ 5\degr (1.0 $\pm 0.1$ rad) we obtain a Mach number 
${\cal M} = \sin(\theta/2)^{-1}$ = 2.1 $\pm 0.5$. This is 
considerably lower than than the ${\cal M} = 9$ value of 
\citet{K+02}, and closer to the theoretical value for a pulsar 
moving through a SNR interior \citep[${\cal M}\approx 3.1$,][]{V+98}.

We can calculate the sound speed of the hot gas within the SNR
according to

\[
c_s = \left( \frac{\gamma k T_e}{\mu m_H} \right)^{\frac{1}{2}}
\]

\noindent where $\gamma$ is the adiabatic index, here taken as 5/3; 
$k$ is Boltzmann's constant;  $T_e$ is the electron temperature;
$\mu$ is the reduced mass; and $m_H$ is the hydrogen particle mass.  
Presuming the reduced mass $\mu = 0.61$, and using 
the temperature found above, we obtain a sound speed in the hot gas 
of 230 $\pm$ 110 km s$^{-1}$. According to ${\cal M} = u/c_s$, where 
$u$ is the motion of the pulsar relative to the SNR material, this gives 
a mean relative motion of $u$=480 $\pm$ 240 km~s$^{-1}$. 

If the pulsar is, as it appears, moving radially outward from the 
SNR center, we can presume its motion to be parallel to that of the
expanding SNR material around the pulsar.  In order to obtain a mean 
relative velocity of $u$=480 km s$^{-1}$ with material moving at the
expansion velocity $v_{\rm exp}$ = 202 km s$^{-1}$ (slow-expansion 
case), we would require the pulsar to be moving at a speed of roughly 
680 km s$^{-1}$ with respect to the SNR center. If instead we use the 
X-ray derived expansion speed of $v_{\rm exp}$ = 470 km s$^{-1}$ 
(fast-expansion case), the pulsar must be moving with a speed of 
$\sim$950 km s$^{-1}$ with respect to the SNR center.   If the motion 
is entirely  perpendicular to the line of sight, and the pulsar began 
at the geometric center of the SNR, it would have taken the pulsar about 
28,000 yr to reach its current transverse distance of 19.4 pc in the 
slow-expansion case, and 20,000 yr in the fast-expansion case.  These 
can, of course, be affected by the viewing angle; but if we assume the 
pulsar to be interacting with the SNR interior, its maximum distance of 
travel is the SNR radius of 21 pc (at a 67\degr\ angle to the line of sight), 
which puts its time of travel at 30,000 yr in the slow-expansion case and 
22,000 yr in the fast-expansion case. 

In our suggested picture, the bow-shocked gas is primarily thermal, 
representing material within the SNR that is shocked to higher temperatures 
by the encounter with the bow shock.  Presumably, the nonthermal X-ray 
emission is associated with the elongated PWN, as would be the radio 
emission. Thus we would expect a broad outer ``cone" of thermal (shocked 
SNR material) emission, with an interior ``cone" of nonthermal (shocked 
PWN) emission.  Such a scenario is in accord with hydrodynamic models of 
bow-shock PWN, in which ``the Mach cone should manifest itself only
in the outer bow shock" \citep[][and references therein]{G+04}. The
shocked PWN material forms a trail of synchrotron emitting particles
opposite the direction of pulsar motion, creating a ``cometary" or 
``linear" morphology.  

To examine this picture for consistency with the data, we fit a simple
thermal plasma plus power-law model combination to data from two regions 
within the linear feature: the thin ``trail" region, corresponding to the 
narrow location of the brightest radio feature (Region 11 in 
Table~\ref{tab:reglist}, Fig.~\ref{fig:xray_regions}f), and the ``bow" 
region, comprised of emission around the linear feature excluding the 
aforementioned ``trail" (Region 10, Fig.~\ref{fig:xray_regions}f).  Both 
regions excluded emission within 6\arcsec\ of the presumed pulsar. The 
best-fit model combination gave thermal and power-law parameters 
similar to those given in previous sections ($N_H\sim 2.6 \times 10^{21}$ 
cm$^{-2}$, $T\sim0.26$ keV, $\Gamma\sim$2.3).  Calculating the fraction of 
the total flux for each model component, we find that the ``trail" region 
shows 96\% nonthermal flux and only 4\% due to thermal plasma, while the 
``bow" region shows 49\% thermal flux and 51\% nonthermal flux.  Errors 
on these flux ratios, based on the differences in flux due to uncertainties 
in the fitted parameters at the 90\% confidence level, are of order 10\%.

Alternately, it is possible that some of the X-ray brightening we 
consider to be associated with the radio ``linear feature" (excluding 
the X-ray knot) is in fact only a surface brightness 
fluctuation, and does not accurately represent the ``Mach cone" of
the presumptive moving pulsar.  If the ``Mach cone" is more tightly 
confined than our X-ray based estimate, the resulting Mach number may 
be somewhat higher, up to a value of about ${\cal M} = 4$ when one 
considers only the angular extent of the bright X-ray emission 
immediately surrounding the X-ray knot.  While it is also
possible that the emission from these features is due to the PWN
alone, with no substantial bow-shock emission, the highly 
elongated morphology and off-center pulsar position argue strongly
for the presence of a significant bow-shock contribution.

\section{SNR age}

The ages of supernova remnants are notoriously ill-characterized.  We
consider a number of different approaches to this age, in order to find
the range of reasonable age estimates.  From the fitted X-ray parameters, 
several estimates are possible.  Combining the derived hot gas density 
with the fitted ionization parameter $\tau$ gives an upper limit age 
estimate  of $\sim$40,000 years for this remnant. However, an age this 
large would be difficult to reconcile with the fact that the hot gas in 
the SNR has apparently not yet reached ionization equilibrium.  Lowering 
the filling factor shortens this age estimate  significantly. A filling 
factor of 0.25, for instance, gives an age estimate of 23,000 yr, similar 
to the estimate based on the radio ``linear feature" by \citet{K+02}.

We can also make another estimate from the Sedov relations, in which the 
relative fractions of thermal and kinetic energies to total energy are 
constant.  Taking the theoretical relation of $E_{\rm th} \approx 0.7 E_0$, 
our estimate for thermal energy (for $f_{\rm hot} = 1$) would yield an initial explosion energy $E_0$ of $8 \times 10^{50}$ erg.  Assuming the density behind 
the shock is a factor of 4 greater than that of the unshocked ISM, we obtain 
an ISM mass density $\rho_0$ of 1.2 $\times 10^{-25}$ g cm$^{-3}$.  Then from 
the Sedov relation $R = 1.17  (E_0/\rho_0)^{\small 1/5}  t^{\small 2/5}$, we 
have an age of only 9,000 yr.  If the X-ray filling factor is lower than 1, 
the age estimate will rise accordingly; $f_{\rm hot} = 0.25$, for instance,
gives an age of 17,000 yr.  If we instead presume the overall ambient density 
to be represented by the current filaments of warm ionized gas, we obtain an 
ISM mass density of 5.3 $\times 10^{-24}$ g cm$^{-3}$. The Sedov relation 
then gives an age as high as 57,000 yr, demonstrating the critical role
played by the ambient density assumption. 

Using the  the simple expansion relation $t=\eta R / v_{\rm exp}$, 
and assuming Sedov expansion ($\eta$=0.4), we can calculate the 
age of the remnant based on the expansion velocity.  The expansion
velocity of 470 km s$^{-1}$ derived from the X-ray temperature gives 
a remnant age of 17,000 yr.  The expansion velocity obtained from 
optical echelle spectroscopy of 202 km s$^{-1}$, however, leads to a
calculated age of 41,000 yr. Again, however, this must 
be considered an upper limit, as this value of $v_{\rm exp}$ is, as 
stated above, a lower limit to the blast-wave expansion.

We can try to narrow down these estimates by considering the 
consistency of the entire picture.  Ages above $\sim$40,000 yr are 
clearly ruled out by the ionization timescale of the X-ray gas and 
by the observed optical expansion.  The value of 9000 yr from 
energetics arguments is highly dependent on the assumption of 
ambient density; a better lower limit might be that of 17,000 yr 
from the expansion velocity derived from the X-ray temperature.
When we consider the estimates of pulsar motion (\S 4.4) we find 
further limitations: the minimum time for the pulsar to reach its 
current location is $\sim$20,000 yr, while the maximum is 
$\sim$30,000 yr.   

If we choose a velocity intermediate between the X-ray and optically 
derived values (e.g., $\sim$300 km s$^{-1}$) and a hot gas filling 
factor of 0.25, we can reconcile most of these numbers.  The ionization
timescale for the hot gas then gives an approximate age of 23,000 yr.
Pulsar travel times give ages (depending on viewing angle) between 
24,000 -- 26,000 yr.  Simple Sedov expansion at this velocity gives
an age of 27,000 yr.  While arguments from Sedov energetics would, 
for this same filling factor, give us a somewhat lower age of 17,000 yr, 
this estimate is highly dependent on the assumption used for ambient 
density, a very uncertain parameter.   We therefore suggest the SNR's
age most likely falls in the range between 23,000 -- 27,000 yr, with 
``hard" limits of 17,000 yr -- 40,000 yr.

\section{Integrated Picture}

We have observed the SNR in N206 with high-resolution optical and X-ray
instruments, and analyzed the results in concert with additional radio 
data and optical echelle spectra.  We find it highly probable that 
N206 is the result of a Type II SN, due to its proximity to other 
massive-star phenomena, enhancement of oxygen abundances, and the 
presence of a probable compact object.

We use these data to calculate overall properties for the physical 
components of the SNR, and find these typical of a middle-aged SNR in 
the adiabatic stage.  The remnant is over-pressured, and the bulk of 
the energy budget still resides in thermal energy from the hot interior.
Evidently, therefore, the hot gas within the SNR still plays a significant
role in the remnant's development.

This remnant is particularly unusual for the very different characteristics 
it displays in different wavelength regimes.  The optical morphology is 
limb-brightened and highly filamentary; the radio morphology is center-filled,
with diffuse emission over the remnant's face, but some SNR shell structure; 
and the X-ray morphology appears somewhat centrally brightened.  The best 
categorization for this remnant appears to be that of ``mixed-morphology" 
\citep{RP98}, but the picture is complicated by the presence of radio and 
X-ray emission near the ``linear feature".  

We analyze the X-ray data for the area surrounding the ``linear feature"
seen in radio. We find a small, hard X-ray source located at the tip of 
the radio feature, with a surface brightness profile consistent with the
presence of an embedded compact source.  Emission from this source and 
its surroundings is nonthermal, with a power-law index similar to that 
seen in Crab-type objects.  The ratio of nonthermal to thermal X-ray flux 
decreases with increasing distance from this source. We conclude that the 
most probable scenario for this feature is a pulsar moving at moderate 
velocity through the surrounding SNR. This creates a bow-shock structure 
in the direction of motion, deforms the surrounding PWN, and leaves behind 
a trail of synchrotron emission along the line of travel.

However, it would be difficult to attribute most, or even the majority, of
the central X-ray emission to that linear feature.  The X-ray emission from 
the central region is clearly dominated by thermal line emission, rather 
than nonthermal continuum.  The lifetime of relativistic particles 
energetic enough to generate X-ray emission is not sufficient for those 
particles to be responsible for the central X-rays, as indicated by the 
fact that the radio emission, generated by longer-lived particles, shows 
no particular increase in this central region. 
This indicates, somewhat surprisingly, that N206 is both a ``composite"
SNR, containing a PWN and shell structure together, and a ``mixed 
morphology" SNR, with centrally concentrated X-ray emission that cannot
be simply associated with the PWN. Indeed, N206 is not the only SNR
to show this combination; for instance, the Galactic remnant
W28 combines centrally concentrated X-ray emission with a probable 
off-center PWN \citep{RB02,K92}.  Other examples include 
W44 \citep{C+99,R+94,HHH96} and IC443 \citep{Kawa+02,BB01,O+01,P+88}.  
We shall call this category ``mixed-composite" SNRs, reflecting the 
combined action of separate physical processes influencing their emission.

\acknowledgements
The authors thank the anonymous referee for a very detailed critique of 
this work, which has much improved it, and also thank Brian D. Fields 
for valuable discussions. RMW and YHC acknowledge support from STSCI 
grant STI GO-08110. RMW, YHC, and JRD acknowledge support from NASA 
grant NAG 5-11159. RMW, YHC, FDS and JRD acknowledge support from 
SAO grant G03-4096. MAG acknowledges support from the grant 
AYA~2002-00376 of the Spanish MCyT (cofunded by FEDER funds).

\clearpage

% ===== Table 1 =====

\begin{deluxetable}{lccccc}
\tablecaption{Optical Echelle Observations}
\tablehead{
\colhead{Date} &
\colhead{Line} &
\colhead{Exposure} &
\colhead{Position} 
}
\startdata
2000 Dec 6 & \ha\ + \ion{N}{2} & 1200 s & E$-$W through ``linear feature"\\
2000 Dec 6 &  \ha\ + \ion{N}{2} & 1200 s & N$-$S  through SNR center \\
2004 Jan 14 &  \ha\ + \ion{N}{2} & 1200 s & E$-$W through bright filament\\
\enddata
\label{tab:echobs}
\end{deluxetable}

\clearpage

% ===== Table 2 =====

\begin{deluxetable}{lccccc}
\tablecaption{X-ray Regions}
\tablehead{
\colhead{Label} &
\colhead{Name} &
\colhead{ACIS cts} &
\colhead{MOS cts} &
\colhead{Hardness} 
}
\startdata
\sidehead{Joint Chandra ACIS / XMM MOS regions}
1 & Whole SNR & 19,950 & 10,920 & $-$0.49 \\
2 & Outer Limb & 7,450 & 4,130 & $-$0.45 \\
3 & Central & 6,350 & 3,320 & $-$0.58 \\
4 & Wedge & 2,375 & 1,400 & $-$0.46 \\
5 & Linear Feature & 1350 & 430 & $-$0.24 \\
\sidehead{Small spatial regions (Chandra ACIS only)}
6 & Knot & 456 & \nodata & 0.26 \\
7 & 22\arcsec\ W of knot & 301 & \nodata & $-$0.35 \\
8 & 44\arcsec\ W of knot & 264 & \nodata & $-$0.43 \\
9 & Point Src & 94 & \nodata & 0.50 \\
10 & Bow Shock & 754 & \nodata &  $-$0.45 \\
11 & Trail & 196  & \nodata & 0.076 \\
\enddata
\label{tab:reglist}
\tablecomments{Background-subtracted source counts.  ``Hardness" refers to the
hardness ratio derived from Chandra ACIS counts. The ``Hard" band is defined as
1.0$-$8.0 keV, the ``Soft" band as 0.3$-$1.0 keV; the hardness ratio is H$-$S/H+S.}
\end{deluxetable}

\clearpage

% ===== Table 3 =====

\begin{deluxetable}{lccccccccccccccccc}
\tablecaption{Fits to X-ray emission for N206 SNR}
\label{tab:snrspecfit}
\tablewidth{6.75in}
\tabletypesize{\footnotesize}
\tablehead{
\colhead{Spectral} &
\colhead{$N_{\rm H}$} &
\colhead{$kT$} &
\colhead{abundance} &
\colhead{$\tau$} &
\colhead{norm} &
\colhead{$\chi^2$} &
\colhead{dof}\\
\colhead{Model} &
\colhead{cm$^{-2}$} &
\colhead{keV} &
\colhead{frac. solar} &
\colhead{cm$^{-3}$ s} &
\colhead{cm$^{-5}$} &
\colhead{reducd} &
\colhead{} 
}
\startdata
\cutinhead{Outer Limb Region (excludes ``linear feature")}
\sidehead{Merged ACIS}
pshock  & 2.7$_{-0.5}^{+1.1}\times10^{21}$ & 0.47$_{-0.11}^{+0.04}$ & 0.25$_{-0.08}^{+0.03}$  
	& 1.4$_{-0.4}^{+0.9} \times10^{11}$ & 7$\pm 5 \times10^{-4}$ & 1.38 & 154 \\
mekal   & 3.6$_{-0.2}^{+0.8}\times10^{21}$ & 0.22$_{-0.03}^{+0.01}$ & 0.14$_{-0.03}^{+0.09}$  
	& \nodata & 6$\pm 2 \times10^{-4}$ & 1.93 & 155 \\ 	
\sidehead{EPIC MOS1\&2}
pshock & 7$_{-3}^{+5}\times10^{20}$ & 0.50$_{-0.08}^{+0.06}$ & 0.20$_{-0.06}^{+0.09}$  
	& 3.4$_{-1.0}^{+1.2} \times10^{11}$ & 8$\pm 2 \times10^{-3}$ & 1.17 & 83 \\
mekal  & 2.3$_{-0.4}^{+0.4}\times10^{21}$ & 0.24$_{-0.01}^{+0.02}$ & 0.08$_{-0.02}^{+0.03}$  
	& \nodata & 0.1$\pm 0.02 $ & 1.48 & 84 \\ 
\sidehead{ACIS + EPIC MOS1\&2}
pshock &  2.2$_{-0.4}^{+0.5}\times10^{21}$ & 0.44$_{-0.06}^{+0.04}$ & 0.22$_{-0.04}^{+0.04}$  
	& 2.1$_{-0.5}^{+0.6} \times10^{11}$ & 7$\pm 3 \times10^{-4}$ & 1.54 & 241 \\
		& & & & & 1.5$\pm 0.3 \times10^{-2}$  \\ 
mekal  & 3.1$_{-0.2}^{+0.3}\times10^{21}$ & 0.23$_{-0.01}^{+0.01}$ & 0.12$_{-0.02}^{+0.05}$  
	& \nodata &  6$\pm 2 \times10^{-3}$  & 1.96 & 242 \\ 
	& & & & & 0.12 $\pm$ 0.03 \\ 
\cutinhead{Central Region (excludes ``linear feature")}
\sidehead{Merged ACIS}
pshock  & 3.5 $\pm 0.4 \times10^{21}$ & 0.34 $\pm$ 0.08 & 0.25 $\pm$ 0.05 
	& 4 $\pm 2 \times10^{11}$ & 1.4 $\pm 0.9 \times10^{-3}$ & 3.09 & 91 \\
mekal   & 3.9 $\pm 0.3 \times10^{21}$ & 0.22 $\pm$ 0.08 & 0.4 $\pm$ 0.2 
	& \nodata & 3 $\pm 2 \times10^{-3}$ & 3.79 & 92 \\
\sidehead{EPIC MOS1\&2}
pshock & 2.0 $\pm 0.4 \times10^{21}$ & 0.38 $\pm$ 0.04 & 0.14 $\pm$ 0.03 
	& 2.7 $\pm 0.9 \times10^{11}$ & 2 $\pm 1 \times10^{-3}$ & 2.10 & 64 \\
mekal  & 2.7 $\pm 0.3 \times10^{21}$ & 0.22 $\pm$ 0.01 & 0.11 $\pm$ 0.02 
	& \nodata & 1.0 $\pm 0.4 \times10^{-2}$ & 2.98 & 65 \\ 	
\sidehead{ACIS + EPIC MOS1\&2}
pshock &   2.9$\pm 0.3 \times10^{21}$ & 0.37$\pm$ 0.02 & 0.25$\pm$ 0.03
	& 3.0$\pm 0.2  \times10^{11}$ & 9$\pm 4 \times10^{-4}$ & 3.21 & 159 \\
		& & & & & 2.1$\pm 0.9 \times10^{-3}$  \\ 
mekal  & 3.6$\pm 0.2 \times10^{21}$ & 0.22$\pm$ 0.06 & 0.4$\pm$ 0.1
	& \nodata & 3$\pm 1 \times10^{-4}$ & 3.97 & 160 \\
		& & & & & 6$\pm 2 \times10^{-3}$  \\ 
\enddata
\label{tab:snrspecfit}
\tablecomments{All spectra cover the range between 0.3-10.0 keV.}
\end{deluxetable}

\clearpage

% ===== Table 4 ===== 

\begin{deluxetable}{cccccccccccccccccc}
\tablecaption{Fits to X-ray Emission for ``Whole SNR" Region of the N206 SNR}
\label{tab:snrspecfit3}
\tablehead{
\colhead{Parameter} &
\multicolumn{4}{c}{Model Fits} \\
\colhead{} &
\colhead{CIE} &
\colhead{NEI} &
\multicolumn{2}{c}{Combined NEI + powerlaw} &
}
\startdata
component & vmekal & vpshock & vpshock & powerlaw  \\[3pt]
$N_{\rm H}$ (cm$^{-2}$) &  2.8$^{+0.1}_{-0.1} \times10^{21}$ 
	& 2.2$^{+0.1}_{-0.1} \times10^{21}$ 
	& \multicolumn{2}{c}{2.2$^{+0.1}_{-0.1} \times10^{21}$}\\[6pt]
$kT$ (keV) & 0.264$^{+0.002}_{-0.004}$  &  0.453$^{+0.005}_{-0.006}$
 	& 0.43$^{+0.01}_{-0.01}$ & \nodata \\[6pt]
 O/O$_\sun$ & 0.36$^{+0.02}_{-0.02}$ & 0.32$^{+0.02}_{-0.01}$
 	& 0.30$^{+0.01}_{-0.01}$ & \nodata \\[6pt]
 Ne/Ne$_\sun$ & 0.24$^{+0.03}_{-0.02}$ & 0.22$^{+0.02}_{-0.01}$
 	& 0.24$^{+0.01}_{-0.01}$ & \nodata \\[6pt]
 Mg/Mg$_\sun$ & 0.66$^{+0.06}_{-0.05}$ & 0.39$^{+0.04}_{-0.03}$
 	& 0.41$^{+0.04}_{-0.03}$ & \nodata \\[6pt]
 Si/Si$_\sun$ & 0.9$^{+0.2}_{-0.2}$ & 0.28$^{+0.07}_{-0.08}$
 	& 0.3$^{+0.1}_{-0.2}$ & \nodata \\[6pt]
 Fe/Fe$_\sun$ & 0.18$^{+0.01}_{-0.01}$ &  0.20$^{+0.01}_{-0.02}$
 	& 0.21$^{+0.02}_{-0.02}$ &  \nodata\\[6pt]
$\tau$ (cm$^{-3}$ s) & \nodata & 3.5$^{+0.2}_{-0.2} \times10^{11}$  
 	& 3.3$^{+0.2}_{-0.2} \times10^{11}$  & \nodata \\[6pt]
$\Gamma$ & \nodata & \nodata & \nodata & 2.2$^{+0.3}_{-0.4}$ &  \\[6pt]
ACIS norm  (cm$^{-5}$)& 5.25$^{+0.07}_{-0.07} \times10^{-3}$ 
	& 1.75$^{+0.03}_{-0.02} \times10^{-3}$
 	&  1.76$^{+0.03}_{-0.02} \times10^{-3}$
 	& 2.3$^{+0.5}_{-0.6} \times10^{-5}$ \\[6pt]
MOS norm  (cm$^{-5}$)& 1.05$^{+0.01}_{-0.02} \times10^{-2}$ 
	& 3.47$^{+0.06}_{-0.06} \times10^{-3}$
 	& 3.59$^{+0.07}_{-0.06} \times10^{-3}$
 	& 7$^{+1}_{-7} \times10^{-6}$ \\[6pt]
$\chi^2_{\rm red}$ & 2.73 & 2.23 & \multicolumn{2}{c}{2.15}\\
dof & 294 & 294 & \multicolumn{2}{c}{292} \\
\enddata
\label{tab:wholesnrspec}
\tablecomments{Results in Tables \ref{tab:wholesnrspec}-\ref{tab:linearspec} 
are from simultaneous fits to ACIS and EPIC MOS data. All spectra cover 
the range between 0.3-8.0 keV. }
\end{deluxetable}

\clearpage

% ===== Table 5 ===== 

\begin{deluxetable}{cccccccccccccccccc}
\tablecaption{Fits to X-ray Emission for ``Outer Limb" Region of the N206 SNR}
\label{tab:snrspecfit3}
\tablehead{
\colhead{Parameter} &
\multicolumn{4}{c}{Model Fits} \\
\colhead{} &
\colhead{CIE} &
\colhead{NEI} &
\multicolumn{2}{c}{Combined NEI + powerlaw} &
}
\startdata
component & vmekal & vpshock & vpshock & powerlaw  \\[3pt]
$N_{\rm H}$ (cm$^{-2}$) &  2.9$^{+0.1}_{-0.2} \times10^{21}$ 
	& 1.7$^{+0.1}_{-0.1} \times10^{21}$ 
	& \multicolumn{2}{c}{1.7$^{+0.1}_{-0.1} \times10^{21}$}\\[6pt]
$kT$ (keV) &  0.237$^{+0.03}_{-0.02}$ & 0.46$^{+0.02}_{-0.01}$
 	& 0.46$^{+0.01}_{-0.01}$  & \nodata \\[6pt]
 O/O$_\sun$ & 0.25$^{+0.04}_{-0.02}$ & 0.22$^{+0.01}_{-0.02}$
 	& 0.22$^{+0.01}_{-0.04}$ & \nodata \\[6pt]
 Ne/Ne$_\sun$ & 0.20$^{+0.03}_{-0.02}$ & 0.25$^{+0.03}_{-0.03}$
 	& 0.25$^{+0.02}_{-0.03}$ & \nodata \\[6pt]
 Mg/Mg$_\sun$ & 0.7$^{+0.1}_{-0.1}$  & 0.43$^{+0.06}_{-0.06}$
 	& 0.43$^{+0.07}_{-0.06}$ &  \nodata\\[6pt]
 Si/Si$_\sun$ & 0.8$^{+0.4}_{-0.3}$ & 0.2$^{+0.2}_{-0.1}$
 	& 0.2$^{+0.2}_{-0.1}$ & \nodata \\[6pt]
 Fe/Fe$_\sun$ & 0.21$^{+0.2}_{-0.3}$ & 0.24$^{+0.02}_{-0.02}$
 	& 0.24$^{+0.02}_{-0.02}$ & \nodata \\[6pt]
$\tau$ (cm$^{-3}$ s) & \nodata & 2.2$^{+0.2}_{-0.2} \times10^{11}$  
 	& 2.2$^{+0.2}_{-0.2} \times10^{11}$  & \nodata \\[6pt]
$\Gamma$ & \nodata & \nodata & \nodata & 2$^{+3}_{-2}$ &  \\[6pt]
ACIS norm  (cm$^{-5}$)& 2.69$^{+0.05}_{-0.06} \times10^{-3}$ 
	&  5.1$^{+0.1}_{-0.1} \times10^{-4}$
 	& 5.0$^{+0.2}_{-0.1} \times10^{-4}$ 
 	& 2$^{+4}_{-2} \times10^{-6}$ \\[6pt]
MOS norm  (cm$^{-5}$)& 5.8$^{+0.2}_{-0.1} \times10^{-2}$ 
	& 1.11$^{+0.02}_{-0.04} \times10^{-2}$
 	&  1.11$^{+0.03}_{-0.04} \times10^{-2}$
 	& 0$^{+2\times10^{-5}}_{-0} $ \\[6pt]
$\chi^2_{\rm red}$ & 1.71 & 1.47 & \multicolumn{2}{c}{1.48}\\
dof & 203 & 203 & \multicolumn{2}{c}{201} \\
\enddata
\label{tab:outerlimbspec}
\end{deluxetable}

\clearpage

% ===== Table 6 ===== 

\begin{deluxetable}{cccccccccccccccccc}
\tablecaption{Fits to X-ray Emission for ``Central" Region of the N206 SNR}
\label{tab:snrspecfit3}
\tablehead{
\colhead{Parameter} &
\multicolumn{4}{c}{Model Fits} \\
\colhead{} &
\colhead{CIE} &
\colhead{NEI} &
\multicolumn{2}{c}{Combined NEI + powerlaw} &
}
\startdata
component & vmekal & vpshock & vpshock & powerlaw  \\[3pt]
$N_{\rm H}$ (cm$^{-2}$) & 2.8$^{+0.01}_{-0.02}  \times10^{21}$ 
	& 3.0$^{+0.1}_{-0.1} \times10^{21}$ 
	& \multicolumn{2}{c}{2.7$^{+0.1}_{-0.2} \times10^{21}$}\\[6pt]
$kT$ (keV) & 0.278$^{+0.002}_{-0.003}$  & 0.335$^{+0.002}_{-0.006}$
 	& 0.37$^{+0.02}_{-0.01}$ & \nodata \\[6pt]
 O/O$_\sun$ & 0.56$^{+0.05}_{-0.04}$  & 0.46$^{+0.02}_{-0.04}$
 	& 0.46$^{+0.04}_{-0.02}$ & \nodata \\[6pt]
 Ne/Ne$_\sun$ & 0.16$^{+0.07}_{-0.03}$ & 0.24$^{+0.03}_{-0.02}$
 	& 0.27$^{+0.04}_{-0.02}$ & \nodata \\[6pt]
 Mg/Mg$_\sun$ & 0.9$^{+0.1}_{-0.1}$ & 0.5$^{+0.2}_{-0.1}$  
 	& 0.6$^{+0.1}_{-0.1}$ & \nodata \\[6pt]
 Si/Si$_\sun$ & 0.7$^{+0.2}_{-0.3}$ & 0.3$^{+0.2}_{-0.2}$
 	& 0.3$^{+0.2}_{-0.2}$ & \nodata \\[6pt]
 Fe/Fe$_\sun$ & 0.22$^{+0.01}_{-0.04}$ &  0.21$^{+0.02}_{-0.03}$
 	& 0.23$^{+0.02}_{-0.02}$ & \nodata \\[6pt]
$\tau$ (cm$^{-3}$ s) & \nodata & 9$^{+1}_{-1} \times10^{11}$  
 	& 5.5$^{+1}_{-1} \times10^{11}$   & \nodata \\[6pt]
$\Gamma$ & \nodata & \nodata & \nodata & 3$^{+3}_{-3}$  \\[6pt]
ACIS norm  (cm$^{-5}$)& 1.22$^{+0.03}_{-0.03} \times10^{-3}$ 
	& 1.13$^{+0.02}_{-0.03} \times10^{-3}$
 	&  7.7$^{+0.2}_{-0.3} \times10^{-4}$
 	& 2$^{+0.3}_{-0.3} \times10^{-6}$ \\[6pt]
MOS norm  (cm$^{-5}$)& 2.72$^{+0.08}_{-0.08} \times10^{-3}$ 
	& 2.5$^{+0.1}_{-0.1} \times10^{-3}$
 	& 1.7$^{+0.1}_{-0.1} \times10^{-3}$
 	& 0$^{+3\times10^{-6}}_{-0}$ \\[6pt]
$\chi^2_{\rm red}$ & 2.50 & 2.22 & \multicolumn{2}{c}{2.18}\\
dof & 150 & 150 & \multicolumn{2}{c}{148} \\
\enddata
\label{tab:centerspec}
\end{deluxetable}

\clearpage

% ===== Table 7 ===== 

\begin{deluxetable}{cccccccccccccccccc}
\tablecaption{Fits to X-ray Emission for ``Wedge" Region of the N206 SNR}
\label{tab:snrspecfit3}
\tablehead{
\colhead{Parameter} &
\multicolumn{4}{c}{Model Fits} \\
\colhead{} &
\colhead{CIE} &
\colhead{NEI} &
\multicolumn{2}{c}{Combined NEI + powerlaw} &
}
\startdata
component & vmekal & vpshock & vpshock & powerlaw  \\[3pt]
$N_{\rm H}$ (cm$^{-2}$) & 2.3$^{+0.2}_{-0.2}  \times10^{21}$ 
	& 2.2$^{+0.1}_{-0.1} \times10^{21}$ 
	& \multicolumn{2}{c}{1.8$^{+0.1}_{-0.1} \times10^{21}$}\\[6pt]
$kT$ (keV) & 0.32$^{+0.01}_{-0.01}$ & 0.90$^{+0.04}_{-0.05}$
 	& 0.66$^{+0.03}_{-0.02}$ & \nodata \\[6pt]
 O/O$_\sun$ & 0.31$^{+0.04}_{-0.04}$ & 0.30$^{+0.03}_{-0.03}$
 	& 0.39$^{+0.04}_{-0.03}$ &  \nodata\\[6pt]
 Ne/Ne$_\sun$ & 0.09$^{+0.02}_{-0.03}$ & 0.12$^{+0.04}_{-0.04}$
 	& 0.14$^{+0.06}_{-0.04}$ & \nodata \\[6pt]
 Mg/Mg$_\sun$ & 0.36$^{+0.08}_{-0.09}$ & 0.2$^{+0.1}_{-0.2}$
 	& 0.4$^{+0.1}_{-0.1}$ & \nodata \\[6pt]
 Si/Si$_\sun$ & 0.8$^{+0.2}_{-0.3}$ &  0.2$^{+0.2}_{-0.1}$
 	& 0.1$^{+0.2}_{-0.1}$ & \nodata \\[6pt]
 Fe/Fe$_\sun$ & 0.06$^{+0.01}_{-0.02}$ &  0.19$^{+0.02}_{-0.03}$
 	& 0.19$^{+0.02}_{-0.04}$ & \nodata \\[6pt]
$\tau$ (cm$^{-3}$ s) & \nodata & 7$^{+1}_{-1} \times10^{10}$  
 	& 2.1$^{+0.1}_{-0.3} \times10^{11}$ & \nodata \\[6pt]
$\Gamma$ & \nodata & \nodata & \nodata & 1.6$^{+0.9}_{-0.9}$ &  \\[6pt]
ACIS norm  (cm$^{-5}$)& 5.9$^{+0.02}_{-0.02} \times10^{-4}$ 
	& 1.05$^{+0.03}_{-0.05} \times10^{-4}$
 	&  1.14$^{+0.05}_{-0.03} \times10^{-4}$
 	&  2$^{+2}_{-1} \times10^{-6}$ \\[6pt]
MOS norm  (cm$^{-5}$)& 4.3$^{+0.2}_{-0.2} \times10^{-2}$ 
	& 7.5$^{+0.3}_{-0.4} \times10^{-3}$
 	&  7.9$^{+0.4}_{-0.4} \times10^{-4}$
 	&  2.3$^{+0.8}_{-0.9} \times10^{-4}$\\[6pt]
$\chi^2_{\rm red}$ & 1.49 & 1.07 & \multicolumn{2}{c}{0.98}\\
dof & 102 & 102 & \multicolumn{2}{c}{100} \\
\enddata
\label{tab:wedgespec}
\end{deluxetable}

\clearpage

% ===== Table 8 ===== 

\begin{deluxetable}{cccccccccccccccccc}
\tablecaption{Fits to X-ray Emission for ``Linear Feature" Region of the N206 SNR}
\label{tab:snrspecfit3}
\tablehead{
\colhead{Parameter} &
\multicolumn{4}{c}{Model Fits} \\
\colhead{} &
\colhead{CIE} &
\colhead{NEI} &
\multicolumn{2}{c}{Combined NEI + powerlaw} &
\colhead{powerlaw}
}
\startdata
component & vmekal & vpshock & vpshock & powerlaw & powerlaw \\[3pt]
$N_{\rm H}$ (cm$^{-2}$) &  1.0$^{+0.4}_{-0.3} \times10^{21}$ 
	& 3.5$^{+0.4}_{-0.5} \times10^{21}$ 
	& \multicolumn{2}{c}{ 2.2$^{+0.5}_{-0.4} \times10^{21}$}
	& 2.6$^{+0.7}_{-0.7} \times10^{21}$\\[6pt]
$kT$ (keV) & 2.0$^{+0.5}_{-0.3}$ & 2.8$^{+0.9}_{-0.6}$
 	& 0.4$^{+0.2}_{-0.1}$ & \nodata & \nodata\\[6pt]
 O/O$_\sun$ & 2$^{+3}_{-2}$ & 0.15$^{+0.06}_{-0.08}$
 	& 0.2$^{+0.4}_{-0.2}$ &  \nodata & \nodata\\[6pt]
 Ne/Ne$_\sun$ & 2$^{+2}_{-2}$ & 0.18$^{+0.07}_{-0.09}$
 	& 0.2$^{+0.5}_{-0.2}$ & \nodata & \nodata\\[6pt]
 Mg/Mg$_\sun$ & 1$^{+2}_{-1}$ & 0.3$^{+0.2}_{-0.1}$
 	& 0.7$^{+0.5}_{-0.5}$ & \nodata & \nodata\\[6pt]
 Si/Si$_\sun$ & 0.1$^{+0.5}_{-0.1}$ & 0.2$^{+0.3}_{-0.2}$
 	& 0.4$^{+0.9}_{-0.4}$ & \nodata & \nodata\\[6pt]
 Fe/Fe$_\sun$ & 0.1$^{+0}_{-0.1}$& 0.11$^{+0.05}_{-0.05}$
 	& 0.1$^{+0.1}_{-0.1}$ & \nodata & \nodata\\[6pt]
$\tau$ (cm$^{-3}$ s) & \nodata &  4$^{+1}_{-2} \times10^{10}$  
 	&  5$\times10^{13}$\tablenotemark{a}  & \nodata & \nodata\\[6pt]
$\Gamma$ & \nodata & \nodata & \nodata & 2.2$^{+0.3}_{-0.3}$ 
	& 2.8$^{+0.2}_{-0.3}$ \\[6pt]
ACIS norm  &  4.5$^{+0.4}_{-0.5} \times10^{-4}$ 
	&  4.2$^{+0.4}_{-0.5} \times10^{-5}$
 	&  3$^{+2}_{-2} \times10^{-5}$
 	& 1.5$^{+0.1}_{-0.3} \times10^{-5}$ 
 	& 2.4$^{+0.7}_{-0.6} \times10^{-5}$ \\[6pt]
MOS norm  &  1.1$^{+0.1}_{-0.2} \times10^{-4}$ 
	& 1.0$^{+0.2}_{-0.1} \times10^{-4}$
 	& 2.0$^{+0.5}_{-0.3} \times10^{-4}$ 
 	& 2.7$^{+0.5}_{-0.7} \times10^{-5}$ 
 	& 6$^{+1}_{-2} \times10^{-5}$ \\[6pt]
$\chi^2_{\rm red}$ & 1.13 & 1.04 & \multicolumn{2}{c}{0.93} & 1.08\\
dof & 92 & 92 & \multicolumn{2}{c}{90} & 92 \\
\enddata
\tablenotetext{a}{Error estimation gave a range of 6$\times10^{12}$ to 
5$\times10^{13}$ cm$^{-3}$ s; the latter is the hard limit for this parameter.
At this upper limit the plasma is considered to be in ionization equilibrium.}
\label{tab:linearspec}
\end{deluxetable}

\clearpage

% ===== Table 9 =====

\begin{deluxetable}{lcccccccc}
\tablecaption{Small-region fits (Chandra ACIS only)}
\tablehead{
\colhead{Region} &
\colhead{kT} &
\colhead{$\tau$} &
\colhead{$\Gamma$} &
\colhead{K$_{\rm pshock}$} &
\colhead{K$_{\rm pow}$} &
\colhead{$\chi^2$/dof}
}
\startdata
 Knot &  0.9$^{+0.8}_{-0.5}$ & 4$^{+8}_{-9} \times10^{10}$ & 1.7$^{+0.2}_{-0.2}$ 
 	& 2.6$^{+0.6}_{-0.9} \times10^{-6}$ & 7$^{+1}_{-1} \times10^{-6}$ & 0.89/18 \\
 22\arcsec\ W of knot & 2.0$^{+0.8}_{-0.7}$ & 3$^{+1}_{-1} \times10^{10}$ 
 	& 2$^{+1}_{-1}$ & 7$^{+1}_{-1} \times10^{-6}$ 
 	& 1.2$^{+0.7}_{-0.7} \times10^{-6}$ & 0.67/15 \\
 44\arcsec\ W of knot &  0.36$^{+0.05}_{-0.04}$ & 8$^{+8}_{-4} \times10^{11}$ 
  	& 3$^{+1}_{-1}$ & 2.4$^{+0.3}_{-0.4} \times10^{-5}$ 
 	& 1.8$^{+0.8}_{-0.9} \times10^{-6}$ & 0.78/15 \\
 Point Source &  \nodata & \nodata & 2.1$^{+0.4}_{-0.5}$ & \nodata 
	&  2.4$^{+0.4}_{-0.5} \times10^{-6}$ & 1.4/6\\
\enddata
\tablecomments{All spectra cover the range between 0.3-8.0 keV. For these fits,
combined NEI plasma (pshock) and power-law (pow) models were used.
N$_H$ was fixed at a value of 2.2$\times10^{21}$ cm$^{-2}$ and abundances were 
set to 30\% solar, consistent with similar fits for this area above. }
\label{tab:smregspec}
\end{deluxetable}

\clearpage

% ===== Table 10 ===== 

\begin{deluxetable}{lccccc}
\tablecaption{Derived Physical Parameters for SNR N206}
\tablehead{
\colhead{Region} &
}
\startdata
Cool shell & density            & 10$\pm$4 cm$^{-3}$ \\
Cool shell & mass               & 160 - 2000 $M_{\sun}$ \\
Cool shell & expansion velocity & 202 $\pm$5 km s$^{-3}$ \\
Cool shell & kinetic energy     & 1 - 8 $\times 10^{50}$ erg \\
Cool shell & thermal pressure   & 3$\pm 1 \times 10^{-11}$ dyne cm$^{-2}$ \\
Hot gas    & temperature        & 0.4$\pm$0.1 keV \\
Hot gas    & density            & 0.24 $f_{hot}^{-\half}$ cm$^{-3}$ \\
Hot gas    & mass               & 270 $\pm$10 $f_{hot}^{\half}$ $M_{\sun}$ \\
Hot gas    & thermal energy     & 6$\pm 1\times 10^{50}$  $f_{hot}^{\half}$ erg \\
Hot gas    & thermal pressure   & 3.3 $\pm 0.4 \times 10^{-10}$ $f_{hot}^{\half}$ dyne cm$^{-2}$ \\
\enddata
\label{tab:parameters}
\end{deluxetable}

\clearpage

% ===== Table 11 =====

\begin{deluxetable}{lccccccccc}
\tablecaption{Linear Feature Thermal vs Nonthermal X-ray Emission (Chandra ACIS only)}
\tablehead{
\colhead{Region} &
\colhead{Power-Law Flux} &
\colhead{Thermal Flux} &
\colhead{Flux Ratio} \\
\colhead{} &
\colhead{erg cm$^{-2}$ s$^{-1}$} &
\colhead{erg cm$^{-2}$ s$^{-1}$} &
\colhead{pwrlaw/thrml} \\ 
}
\startdata
\cutinhead{By fitted spectral model components}
 Knot & 2.2$\pm 0.6 \times 10^{-14}$ & 2$\pm 1 \times 10^{-15}$ & 0.9/0.1 \\
 22\arcsec\ W of knot & 7$\pm 1\times 10^{-15}$ & 7.5$\pm 0.5 \times 10^{-15}$ & 0.5/0.5 \\
 44\arcsec\ W of knot & 2.7$\pm 0.4\times 10^{-15}$ & 7.9$\pm 0.7 \times 10^{-15}$ & 0.3/0.7 \\
\cutinhead{By energy range (Thermal: 0.3-1.0 keV; Powerlaw: 2.0-8.0 keV)}
 Knot & 2.3$\pm 0.2 \times 10^{-14}$ & 4.3$\pm 0.3 \times 10^{-15}$ & 0.8/0.2 \\
 22\arcsec\ W of knot & 8.3$\pm 0.1 \times 10^{-15}$ & 7.3$\pm 0.2 \times 10^{-15}$ & 0.5/0.5 \\
 44\arcsec\ W of knot & 8$\pm 1 \times 10^{-16}$ & 7.3$\pm 0.1 \times 10^{-15}$ & 0.1/0.9 \\
\enddata
\label{tab:linearflux}
\end{deluxetable}

\clearpage

\clearpage

\clearpage

\clearpage

\clearpage

\clearpage

\clearpage

\end{document}